\documentclass[twocolumn]{aastex631}

\usepackage{amsmath,amstext}
\usepackage{gensymb}
\usepackage{subfigure}
\usepackage[T1]{fontenc}
\usepackage{xcolor}

\shorttitle{Sgr E}
\shortauthors{Wallace et al.}
\begin{document}

\title{ALMA uncovers highly filamentary structure towards the Sgr E region}

\correspondingauthor{J. Wallace}
\email{jennifer.2.wallace@uconn.edu}

\author{J. Wallace}
\affiliation{University of Connecticut, Department of Physics, 196A Auditorium Road Unit 3046, Storrs, CT 06269 USA}

\author{C. Battersby}
\affiliation{University of Connecticut, Department of Physics, 196A Auditorium Road Unit 3046, Storrs, CT 06269 USA}

\author{E. A. C. Mills}
\affiliation{Department of Physics and Astronomy, University of Kansas, 1251 Wescoe Hall Dr., Lawrence, KS 66045 USA}

\author{J. D. Henshaw}
\affiliation{Astrophysics Research Institute, Liverpool John Moores University, IC2, Liverpool Science Park, 146 Brownlow Hill, Liverpool L3 5RF, UK}
\affiliation{Max Planck Institut f\"{u}r Astronomie, K\"{o}nigstuhl 17, D-69117 Heidelberg, DE}

\author{M. C. Sormani}
\affiliation{Universit\"at Heidelberg, Zentrum f\"ur Astronomie, Institut f\"ur Theoretische Astrophysik, Albert-Ueberle-Str. 2, D-69120 Heidelberg, Germany}

\author{A. Ginsburg}
\affiliation{University of Florida, Department of Astronomy, Bryant Space Science Center, Gainesville, FL 32611 USA}

\author{A. T. Barnes}
\affiliation{Argelander-Institut f\"ur Astronomie, Universit\"at Bonn, Auf dem H\"ugel 71, 53121, Bonn, DE}

\author{H. P. Hatchfield}
\affiliation{University of Connecticut, Department of Physics, 196A Auditorium Road Unit 3046, Storrs, CT 06269 USA}

\author{S. C. O. Glover}
\affiliation{Universit\"at Heidelberg, Zentrum f\"ur Astronomie, Institut f\"ur Theoretische Astrophysik, Albert-Ueberle-Str. 2, D-69120 Heidelberg, Germany}

\author{L. D. Anderson}
\affiliation{Department of Physics and Astronomy, West Virginia University, Morgantown, WV 26506, USA}



\begin{abstract}

We report on the discovery of linear filaments observed in CO(1-0) emission for a $\sim2\arcmin$ field of view toward the Sgr E star forming region centered at $(l,b)=(358.720\degree, 0.011\degree$). The Sgr E region is thought to be at the turbulent intersection of the ``far dust lane'' associated with the Galactic bar and the Central Molecular Zone (CMZ). This region is subject to strong accelerations which are generally thought to inhibit star formation, yet Sgr E contains a large number of HII regions.  We present $^{12}$CO(1-0), $^{13}$CO(1-0), and C$^{18}$O(1-0) spectral line observations from ALMA and provide measurements of the physical and kinematic properties for two of the brightest filaments. These filaments have widths (FWHM) of $\sim0.1$ pc and are oriented nearly parallel to the Galactic plane, with angles from the Galactic plane of $\sim2\degree$. The filaments are elongated, with lower limit aspect ratios of $\sim$5:1. For both filaments we detect two distinct velocity components that are separated by about 15 km~s$^{-1}$. In the C$^{18}$O spectral line data with $\sim$0.09 pc spatial resolution,  we find that these velocity components have relatively narrow ($\sim$1-2~km~s$^{-1}$) FWHM linewidths when compared to other sources towards the Galactic center. The properties of these filaments suggest that the gas in the Sgr E complex is being ``stretched'' as it is rapidly accelerated by the gravitational field of the Galactic bar while falling towards the CMZ, a result that could provide insight into the extreme environment surrounding this region and the large-scale processes which fuel this environment.

\end{abstract}


\section{Introduction} 
\label{section:intro}

The Central Molecular Zone (CMZ), spans the innermost radial 250 pc of the Galaxy and contains roughly 5\% of the Galaxy's total molecular gas content \citep{morris_serabyn_1996, Dahmen_1998}. The CMZ is one of the most active star forming regions within the Galaxy, yet it is underproducing stars given the amount of dense gas it has \citep{Longmore_2013,Barnes_2017}. In comparison to molecular clouds found in the Galactic disk, those observed in the CMZ are characterized by higher densities \citep{Mills_2018} and temperatures \citep[e.g.][]{Mills_2013, Ginsburg_2016} as well as larger velocity dispersions \citep[e.g.][]{Shetty_2012, Henshaw_2016}. These properties offer a unique opportunity to observe conditions that may be analogous to those that occur in high redshift galaxies, but at a much closer distance \citep{kru_longmore_2013}.

The conditions observed in the CMZ are thought to be partly maintained by the constant transport of material from the disk of the Galaxy towards the center. This inflow of gas into the CMZ is driven by the Galactic Bar \citep[e.g.][]{Tress_2020}, fueling active star formation and possibly contributing to the high turbulence of the CMZ  \citep{Kruijssen_2014,Federrath_2016,Sormani_Barnes_2019,Salas_2021}. The gas flowing in the gravitational field of the Galactic bar settles into two families of stable orbits, $x_1$ and $x_2$ orbits. The $x_1$ orbits are elongated along the bar's major axis, while $x_2$ orbits exist within the innermost few hundred pc of the Galaxy and are elongated perpendicular to the bar's major axis. At certain energies, the innermost $x_1$ orbits become self-intersecting, causing the gas along the orbit to experience shocks and collisions which cause them to lose angular momentum and settle into the deeper $x_2$-like orbits. The infall of gas from $x_1$ orbits to $x_2$ orbits physically manifests as streams of gas that flow from a Galactocentric radius of R$_{\rm gal}$ = 3 kpc into the CMZ along the so-called Galactic ``dust lanes" \citep{Sormani_Barnes_2019}. These central bar dust lanes are known as the `near' and `far' dust lanes because they lie on opposite sides of the Galactic center, whose major axis forms an angle of $\sim$ 15-30$\degree$ with respect to the Sun-Galactic center line \citep{Bland-Hawthorn_Gerhard_2016}. Although the location where the gas along the dust lanes intersects with the CMZ is not exactly known, there is some observational evidence for interaction between the dust lanes and the CMZ at distance of $\sim$100-200 pc from the Galactic center, specifically towards the Sgr E and 1.3 degree cloud complexes \citep[][and see references therein]{Henshaw_2022}.
The intersection regions between the dust lanes and the CMZ host some of the highest line-of-sight gas velocities ($v_{\rm LSR} \approx$ 270 km s$^{-1}$ and $v_{\rm LSR} \approx$ -220 km s$^{-1}$) observed in the entire Milky Way disk \citep{Dame_2001}. Recent hydrodynamical simulations suggest that approximately one-third of this gas accretes onto the CMZ at a rate of $\sim$ 1 $M_{\odot}$ yr$^{-1}$ \cite{Sormani_Barnes_2019}, while the remaining gas overshoots it \citep{Hatchfield_2021}. Additionally, these intersections between the CMZ and the central bar dust lanes are regions where extreme collisions often occur, resulting in ``extended velocity features'' (EVFs) with extreme velocity dispersions \citep{Sormani_2019}. Therefore, we see that this is a highly dynamic region where turbulent processes occur.

\begin{figure}[htb!] 
\centering
\includegraphics[scale=0.4]{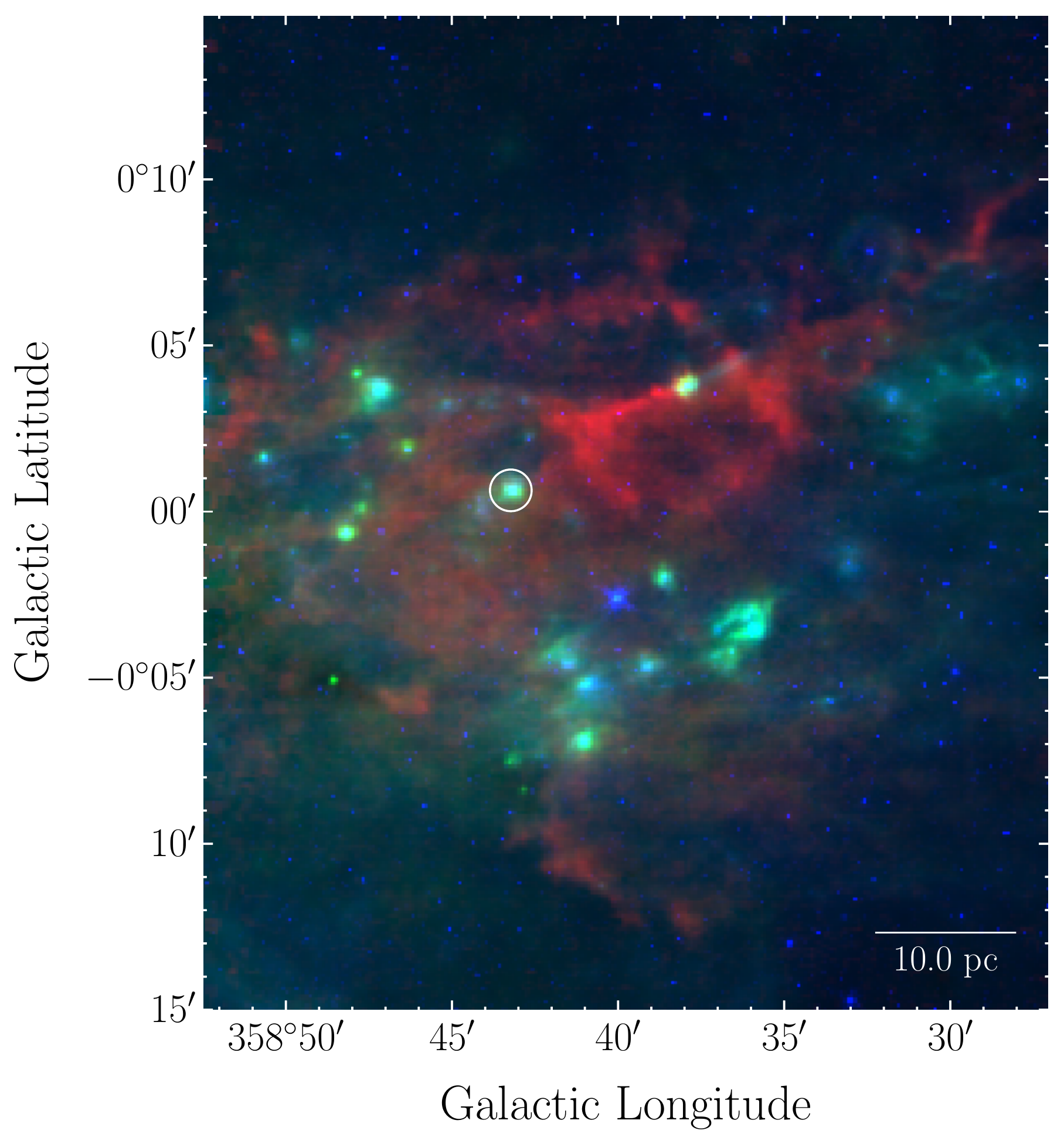}
\caption{
An overview of the Sgr E region. Red is 870 $\mu$m $^{12}$CO (J=3-2) emission integrated over the entire line-of-sight velocity range of $\pm$ 300 km s$^{-1}$ \citep[CHIMPS2;][]{Eden_2020}, green is 70 $\mu$m \citep[Herschel Hi-GAL;][]{Molinari_2010}, and blue is 8 $\mu$m \citep[Spitzer GLIMPSE;][]{Benjamin_2003}. The white circle is the field of view of the ALMA observation. }
\label{fig:overview_rgb}
\end{figure}

\begin{figure}[htb!] 
\centering
\includegraphics[width=0.48\textwidth]{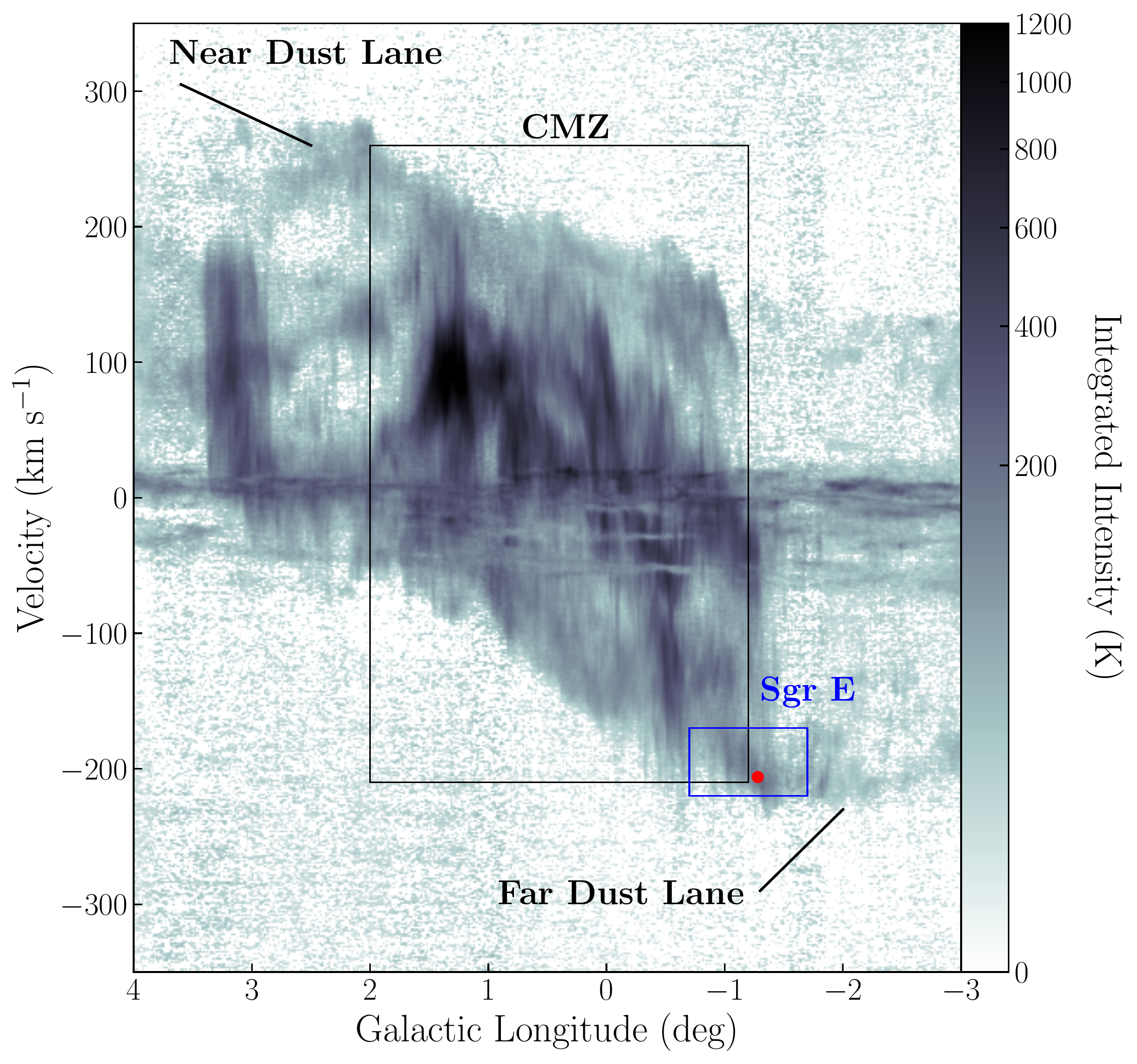}
\caption{An overview of the inner Galaxy in longitude-velocity space as seen in $^{12}$CO (J=3-2) emission \citep[CHIMPS2;][]{Eden_2020}. The black box indicates the longitude range of the CMZ (\textit{l} = 2\degree to -1.1\degree). The blue box represents the line-of-sight velocity range ( -170 km s$^{-1} > v_{\text{LOS}} > -220$ km s$^{-1}$) and longitude range (-0.7\degree $>$ \textit{l} $>$ -1.7\degree) of the Sgr E region. The red circle represents the (\textit{l},\textit{v}) coordinate of the HII region that our data is centered on.} 
\label{fig:overview_lv_full}
\end{figure}

The Sgr E region, seen in Figure \ref{fig:overview_rgb}, is a star forming complex located just outside of the CMZ, spanning Galactic longitudes from \textit{l} $\approx$ 359.3\degree -- 358.3\degree. The placement of the Sgr E complex in both position and line-of-sight velocity is consistent with the intersection point between the dust lane and the CMZ, as seen in Figure \ref{fig:overview_lv_full}. Sgr E has a high negative line-of-sight velocity of $\sim$-200 km s$^{-1}$, consistent with the velocity of gas in the region connecting the far dust lane and the CMZ \citep{Cram_1996, Sormani_Barnes_2019}. Although the Sgr E complex is located near the CMZ, which has a low star formation rate given the amount of dense gas it has \citep{Longmore_2013, Barnes_2017}, it still contains tens of discrete HII regions. We adopt a distance of 8.178 kpc to the Galactic center \citep{grav_collab_2019}, from which we estimate the Sgr E region to be at a projected distance from the Galactic center of $\sim$ 100 - 250 pc. For the purposes of our calculations, we adopt a line-of-sight distance to the Sgr E region of 8.2 kpc.

The radio continuum emission from these regions has been observed and reported on in the literature over the last few decades \citep{Liszt_1992,Gray_1993,Cram_1996,Anderson_2020}, revealing the peculiar characteristics of Sgr E when compared to other Galactic HII region complexes. Unlike the HII regions in other star-forming complexes, the Sgr E HII regions are similar in size and lack a central concentration, with most sources having diameters of 1-4 pc and a median separation of $\sim$ 20 pc between them  \citep{Anderson_2020}. The complex is also associated with a $3 \times 10^{5} ~ M_{\odot}$  molecular cloud that exhibits a strong velocity gradient (-170 km s$^{-1}$ to -220 km s$^{-1}$) from east to west along the Galactic plane \citep{Anderson_2020}.

With a stellar population that is thought to  contain mostly B2 to O8 type stars, Sgr E has an estimated age of 3-5 Myr \citep{Gray_1993,Anderson_2020}. Recent simulations suggest that stars currently contained in the Sgr E complex formed in the far dust lane a few Myr ago and that they will eventually overshoot the CMZ and collide with the near dust lane \citep{Anderson_2020}. 

Observations of the Sgr E complex present a unique opportunity to study a highly dynamic region, with properties that are distinct from other star-forming regions observed in both the CMZ and the Galactic disk. It is anomalous as it is the only star-forming region located at the intersection between the dust lanes and the CMZ. Additionally, the large velocity gradient across the Sgr E region implies that the gas is subject to very strong accelerations. Despite its unusual properties, there are only a handful of papers that focus on the Sgr E complex. It is therefore important to study the Sgr E region in greater detail to gain insight on the properties of the interstellar medium and the process of star formation in highly dynamic environments.

In this paper, we present the first high-resolution ALMA observations in several CO (1-0) emission lines towards a known Sgr E HII region previously identified in \cite{Anderson_2020} located at $l=$ 358.720\degree and $b=$ 0.011\degree, a projected distance of 175 pc from the Galactic center. The HII region has a measured central velocity of -206.1 km s$^{-1}$ and a diameter of 2.7 pc. Our paper is organised as follows: in Section \ref{section:observations} we provide details on the ALMA observations and our final data products. In Section \ref{section:results} we present CO spectral line observations of filaments in the Sgr E region and report on their physical and kinematic structure. In Section \ref{section:discussion} we contextualize the unique properties of these filaments within the broader context of large-scale Galaxy dynamics. Finally, Section \ref{section:conclusion} is a brief summary of our conclusions.

\section{Observations}
\label{section:observations}

\subsection{ALMA data}
\label{section:alma_data}
The data we present in this paper are part of a larger survey observed using the Atacama Large Millimeter/submillimeter Array (ALMA) in Cycle 7 (Project code 2019.1.01240.S, PI: E.A.C. Mills). The observations were completed in 8 sessions with an average of 45 antennas between October 31 and November 24, 2019.  This survey consisted of 25 pointings within the central 5.0 degrees of the Galaxy. In this paper, we present results from a single pointing with a field of view (half power beam width) of 52" (2 pc) toward the Sgr E complex, centered at $l=$358.720\degree and $b=$0.011\degree ($\alpha$=$17^h 42^m 29.90^s$, $\delta$=$-30\degree 1' 14.0''$). The observations were made in the C43-2 configuration, with baselines ranging between 15-697 m.

% Table of cube properties:
\begin{table*}[htb!] 
\centering
\caption{Properties of the CO spectral line data cubes.}

\begin{tabular}{ c c c c c c c c c}
\hline\hline
Spectral Line & Rest Frequency & Beam Position Angle & \multicolumn{2}{c}{Beam Size} &  \multicolumn{2}{c}{LAS\footnote{Largest angular scale, given by $0.6\times \lambda / {b_{\rm min}}$, where $b_{\rm min}$ is the length of the shortest baseline (\url{https://almascience.nrao.edu/about-alma/alma-basics}).}} & \multicolumn{2}{c}{RMS} \\
 & (GHz) & (deg) & (arcsec)  & (pc) & (arcsec)  & (pc) & (mK) & (mJy/beam) \\ 
\hline
$^{12}$C$^{16}$O (1-0) & 115.271202 & -21.3 & $2\arcsec.4\times1\arcsec.4$ & $0.09\times0.06$ & $21\arcsec.2$ & 0.84 & 395 & 13.3 \\
%$^{12}$C$^{17}$O (1-0) & 112.359278 & $2\arcsec.4\times1\arcsec.4$ & $0.09\times0.06$ \\
$^{13}$C$^{16}$O (1-0) & 110.201354 & -28.6 & $2\arcsec.4\times1\arcsec.7$ & $0.09\times0.07$ & $22\arcsec.4$ & 0.89 & 297 & 11.6 \\
$^{12}$C$^{18}$O (1-0) & 109.782176 & -28.6 & $2\arcsec.4\times1\arcsec.7$ & $0.09\times0.07$ & $22\arcsec.6$ & 0.90 & 278 & 10.7\\
\hline
\end{tabular}
\label{table:obs_prop}
\end{table*}

Our observations were made in a single frequency setting at 3 mm (ALMA Band 3) in a total of eight spectral windows. Five of these had bandwidths of 234.38 MHz ($\sim$ 640 km s$^{-1}$\footnote{with respect to the C$^{18}$O spectral line.}) centered on spectral lines, while the remaining three had bandwidths of 1875 MHz covering continuum emission. The continuum spectral windows were centered on the frequencies 97.980953 GHz, 100.900000 GHz, 102.800000 GHz. The remaining spectral windows were centered on spectral lines $^{12}$C$^{16}$O (1-0) (hereafter $^{12}$CO), $^{12}$C$^{17}$O (1-0) (hereafter C$^{17}$O), $^{13}$C$^{16}$O (1-0) (hereafter $^{13}$CO), $^{12}$C$^{18}$O (1-0) (hereafter C$^{18}$O), and H(40)$\alpha$. 

We focus our analysis on the $^{12}$CO, $^{13}$CO, and C$^{18}$O lines. The properties for these spectral line data cubes are given in Table \ref{table:obs_prop}. The calibration of the data was performed in the Common Astronomy Software Applications package (CASA: version 5.6.1-8) using the ALMA pipeline \citep{McMullin_2007}. Imaging of the continuum, as well as the $^{12}$CO, $^{13}$CO, and the C$^{18}$O lines, was performed in CASA using the tclean task, with a robust weighting of 1.0 so that the observations were sensitive to a combination of both point source emission and extended emission. 

The rest frequency, beam size, beam position angle, and per-channel RMS noise for each CO data cube are given in Table \ref{table:obs_prop}. The final $^{12}$CO images have $0\arcsec.28$ pixels (0.011 pc), and the final $^{13}$CO and C$^{18}$O images have $0\arcsec.34$ pixels (0.013 pc). The velocity resolution of the spectral line data is approximately $\sim$ 0.3 km s$^{-1}$ ($\sim$ 0.11 MHz). All images are sensitive to size scales up to 21$\arcsec$ ($\sim$ 0.84 pc) (see LAS in Table \ref{table:obs_prop}).

Before any analyses were performed, we used the
\verb|reproject| and \verb|spectral_interpolate| functions from the Spectral Cube package \citep{adam_ginsburg_2019_3558614} to regrid each data cube to match the C$^{18}$O data cube. The resulting cubes have the same voxel\footnote{A voxel is the unit for a three-dimensional data set, with two axes corresponding to spatial coordinates, in this case Galactic, and the third being a frequency or velocity axis.} sizes, with 0.013 pc pixels and 0.33 km s$^{-1}$ velocity channel widths. For the creation of the line ratio maps, we additionally smooth the cubes to a common beam size taken from the C$^{18}$O data cube.

\subsection{Ancillary data}
\label{subsection:ancillary_data}

In Sections \ref{section:intro} and \ref{section:discussion}, we use $^{12}$CO (J=3-2) spectral line data from the CO Heterodyne Inner Milky Way Plane Survey 2 \citep[CHIMPS2;][]{Eden_2020} to generate longitude-velocity diagrams of the inner Galaxy. The CHIMPS2 observations were completed using the James Clerk Maxwell Telescope (JCMT) and have an angular resolution of 15\arcsec~ and a velocity resolution of 1 km s$^{-1}$.

\section{Results and Analysis}
\label{section:results}

 We investigate the spatial and kinematic properties of the molecular gas in the $^{12}$CO, $^{13}$CO, and the C$^{18}$O spectral line cubes. The bulk of our analysis is performed with the $^{13}$CO and C$^{18}$O data cubes, where we use the brighter $^{13}$CO line emission for filament identification, and the less optically thick C$^{18}$O line emission for measuring the physical properties of each filament. We discuss the relative optical depth for each data cube in Section \ref{subsection:opacity}. Full-resolution channel map images from the $^{12}$CO and C$^{18}$O data cubes are included in Appendix \ref{section:appendix_channels}. 

% Mini Moment Maps: -200 km s$^{-1}$ to -225 km s$^{-1}$
\begin{figure*}[htb!] 
\epsscale{1.15}
\begin{centering}
\plotone{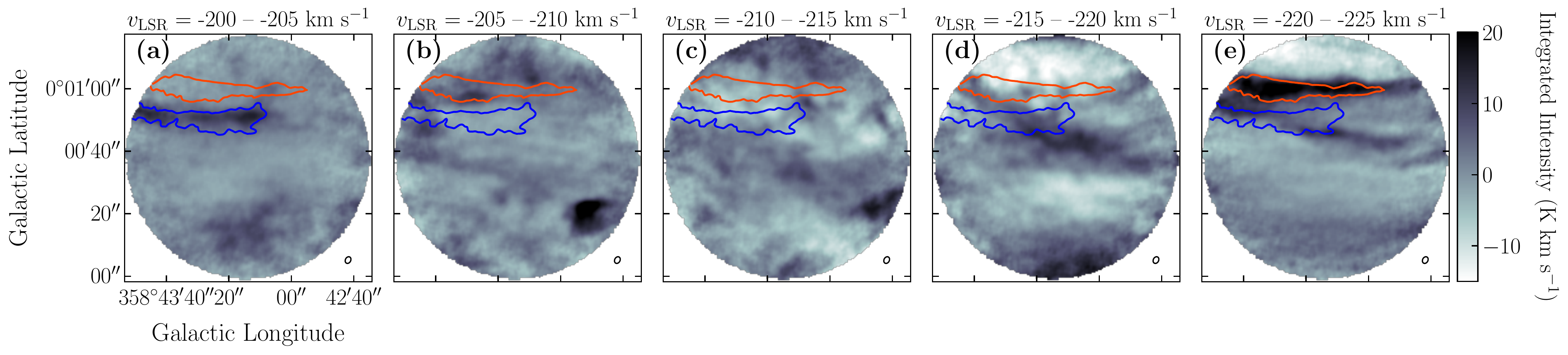}
\caption{Integrated intensity (moment 0) maps from the $^{13}$CO data cube. Each image shows the integrated intensity over a range of 5 km s$^{-1}$. The specific velocity ranges are indicated above each image. The top, orange contour indicates the region we define for filament 1, while the bottom, blue contour indicates the filament 2 region. The beam is indicated in the lower right corner of each channel map. }
\label{fig:mini_mom0}
\end{centering}
\end{figure*}

\subsection{Overview of filamentary structure}
\label{subsection:overview_of_filaments}

We report on a collection of filaments observed in CO emission towards our Sgr E pointing. Our findings indicate the existence of numerous linear filamentary structures that are oriented parallel to the Galactic plane. In this paper, we focus our in-depth analysis on two of the filaments identified (see subsection \ref{subsubsection:fil_ID}).

To provide a brief overview of these filaments, we present five integrated intensity maps from the $^{13}$CO data cube with emission summed over 5 km s$^{-1}$ velocity ranges (Figure \ref{fig:mini_mom0}). These span the overall velocity range of -200 km s$^{-1}$ to -225 km s$^{-1}$. Molecular gas filaments are clearly observed in panels a, d, and e of Figure \ref{fig:mini_mom0}. The orange and blue contours in Figure \ref{fig:mini_mom0} indicate the emission associated with two of the most prominent filaments identified in our data, which we later refer to as ``filament 1" and ``filament 2", respectively. We expand on these filament regions in more detail in Section \ref{subsection:fil_properties}.

In panel a of Figure \ref{fig:mini_mom0}, we identify a single filament, and in Figure \ref{fig:mini_mom0}e, we identify two filaments. All three filaments are very straight and are oriented nearly parallel to the Galactic plane (see Section \ref{subsubsection:phys_char} for more details on the physical characteristics of the filaments). We note that the two filaments observed in Figure \ref{fig:mini_mom0}e are separated by $\sim$ 0.1 pc. The brightest filament in \ref{fig:mini_mom0}e located at \textit{b} = 00\degree 01\arcmin 00\arcsec, has a position angle of approximately $\sim 2 \degree$ with respect to the Galactic plane, whereas the less bright filament has a more pronounced angle of $\sim15\degree$. The less bright filament appears to be co-spatial with the linear structure observed in \ref{fig:mini_mom0}d, which might indicate that the emission observed in both velocity ranges are part of the same filamentary structure.

\subsubsection{Filament Identification}
\label{subsubsection:fil_ID} 

We focus our subsequent analysis on the quantitative properties of a subset of the observed filamentary structures. This subset contains two filaments, which we label filament 1 and filament 2. Both of these filament regions are shown in Figure \ref{fig:mini_mom0} as orange (filament 1) and blue (filament 2) contours. These regions were chosen based on the criterion that they were both observed in emission for the three brightest CO isotopologues. Additionally, filaments 1 and 2 are well-separated from each other, by at least 0.1 pc in position-position space and by more than 10 km s$^{-1}$ in velocity space.

Emission located in the filament 1 region is mostly observed between $v_{\rm LSR}$ = -220 km s$^{-1}$ and -225 km s$^{-1}$ (see Figure \ref{fig:mini_mom0}e). Conversely, in the filament 2 region, the brightest emission is observed between $v_{\rm LSR}$ -200 km s$^{-1}$ and -205 km s$^{-1}$ (see Figure \ref{fig:mini_mom0}a). Another emission feature is seen in Figure \ref{fig:mini_mom0} panels d and e. We omit this feature from our analysis since it is not well separated from the other two filament regions and it is not visible in the C$^{18}$O data. From these integrated intensity maps, we see that filament 1 is located at the Galactic latitude  \textit{b} = 00\degree 01\arcmin 00\arcsec and filament 2 is located at \textit{b} $\sim$ 00\degree 00\arcmin 50\arcsec.

\begin{figure*}[htb!] 
\epsscale{1.15}
\begin{centering}
\subfigure{
\plotone{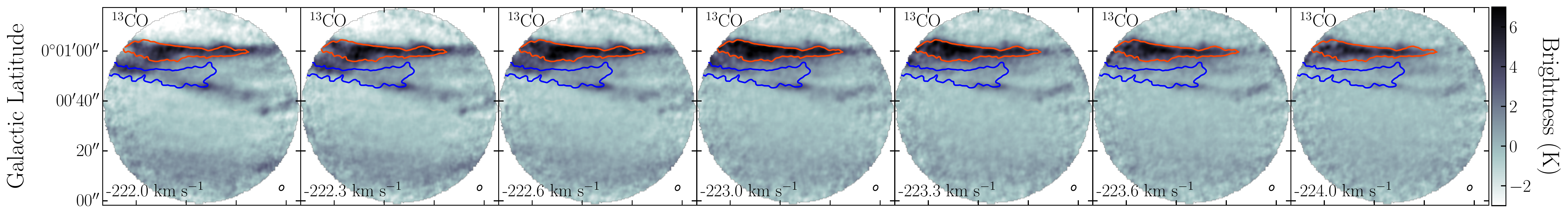}}
\subfigure{
\plotone{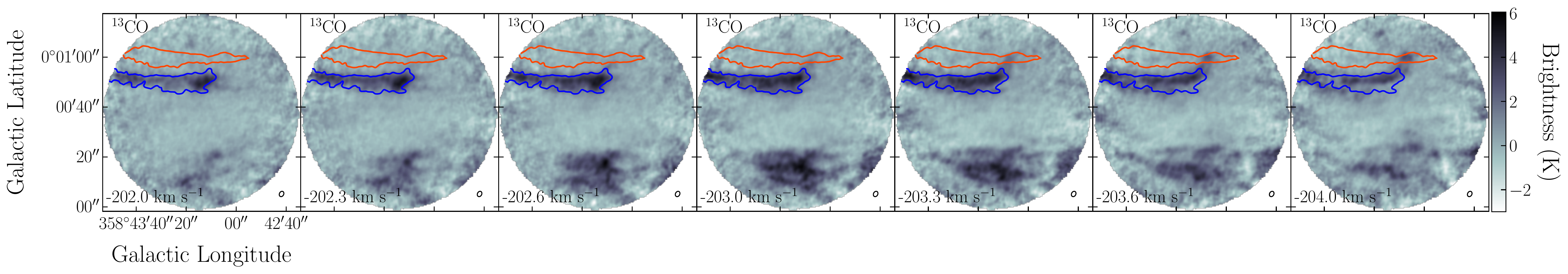}}
\caption{The full resolution velocity channel maps for the spectral line $^{13}$CO used to identify filaments 1 and 2 with channel widths of $\sim$ 0.3 km s$^{-1}$. The channel maps on the top row span the velocity range of -222.0 km s$^{-1}$ to -224.0 km s$^{-1}$. The channel maps on the bottom row span the velocity range of -202.0 km s$^{-1}$ to -204.0 km s$^{-1}$. In each figure, the top, orange contour indicates the region we define for filament 1, while the bottom, blue contour indicates the filament 2 region. The beam is indicated in the lower right corner of each channel map. }
\label{fig:chan13_fil}
\end{centering}
\end{figure*}

We identified these filament regions using full resolution velocity channel maps from the $^{13}$CO data cube. As seen in Figure \ref{fig:chan13_fil}, these channel maps have a channel width of 0.3 km s$^{-1}$ and are centered on the velocity ranges containing the filaments of particular interest. We used the -223 km s$^{-1}$ and -203 km s$^{-1}$ channels of the $^{13}$CO data cube to identify filament 1 and filament 2, respectively (see Figure \ref{fig:chan13_fil}). These two channels were used for quantitative analysis since the emission related to the filaments was brightest at these velocities. We used the $^{13}$CO data for filament identification since it exhibits a higher signal-to-noise compared to the C$^{18}$O data.

% 13CO Velocity channel maps centered on filaments 1 and 2:

To define the filament regions, we created boolean masks from the aforementioned channel map images using a brightness threshold of 4.0 K for the $^{13}$CO -223 km s$^{-1}$ channel and 2.0 K for the $^{13}$CO -203 km s$^{-1}$ channel. These thresholds were chosen by-eye, with the aim of ensuring that the boundaries of the filament regions contained the bulk of the filament emission. To completely isolate each filament, we manually masked out bright regions separated from the filament by more than 0.1 pc, a separation slightly larger than the semimajor axis of our beam ($\sim$ 0.09 pc). 

For velocities spanning -222 km s$^{-1}$ to -224 km s$^{-1}$, the relative shape and brightness of filament 1 varies minimally (Figure \ref{fig:chan13_fil}). We note that the filament below it passes through the filament 2 region at an angle. We also  highlight emission from the filament 2 region in Figure \ref{fig:chan13_fil} for the velocities spanning -202 km s$^{-1}$ to -204 km s$^{-1}$. We note that filament 2 is spatially isolated from other emission structures in the field of view.

\subsubsection{Moment analysis}
\label{subsubsection:moment_analysis}

\begin{figure}[htb!] 
\epsscale{1.0}
\centering
\subfigure{
\plotone{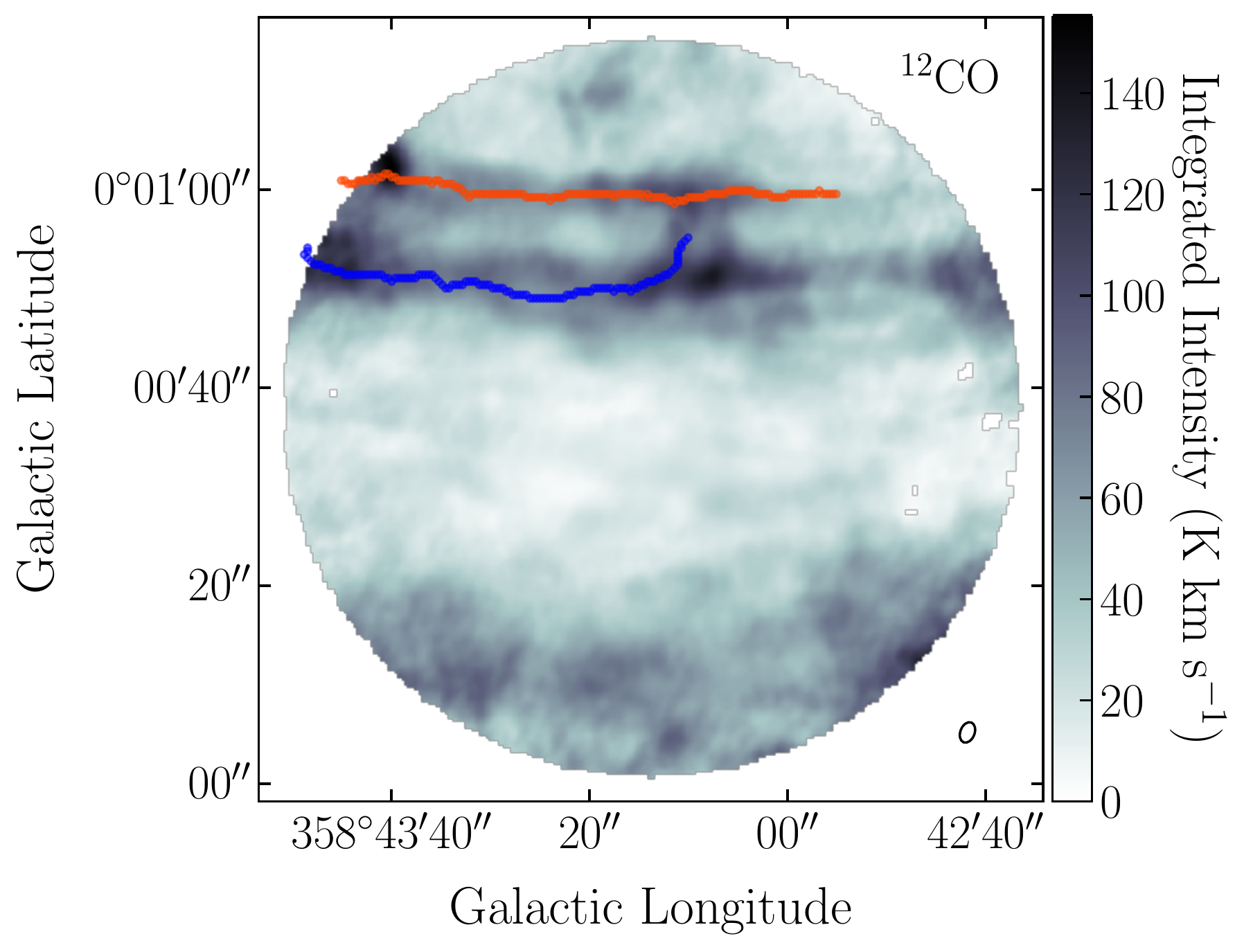}}
\subfigure{
\plotone{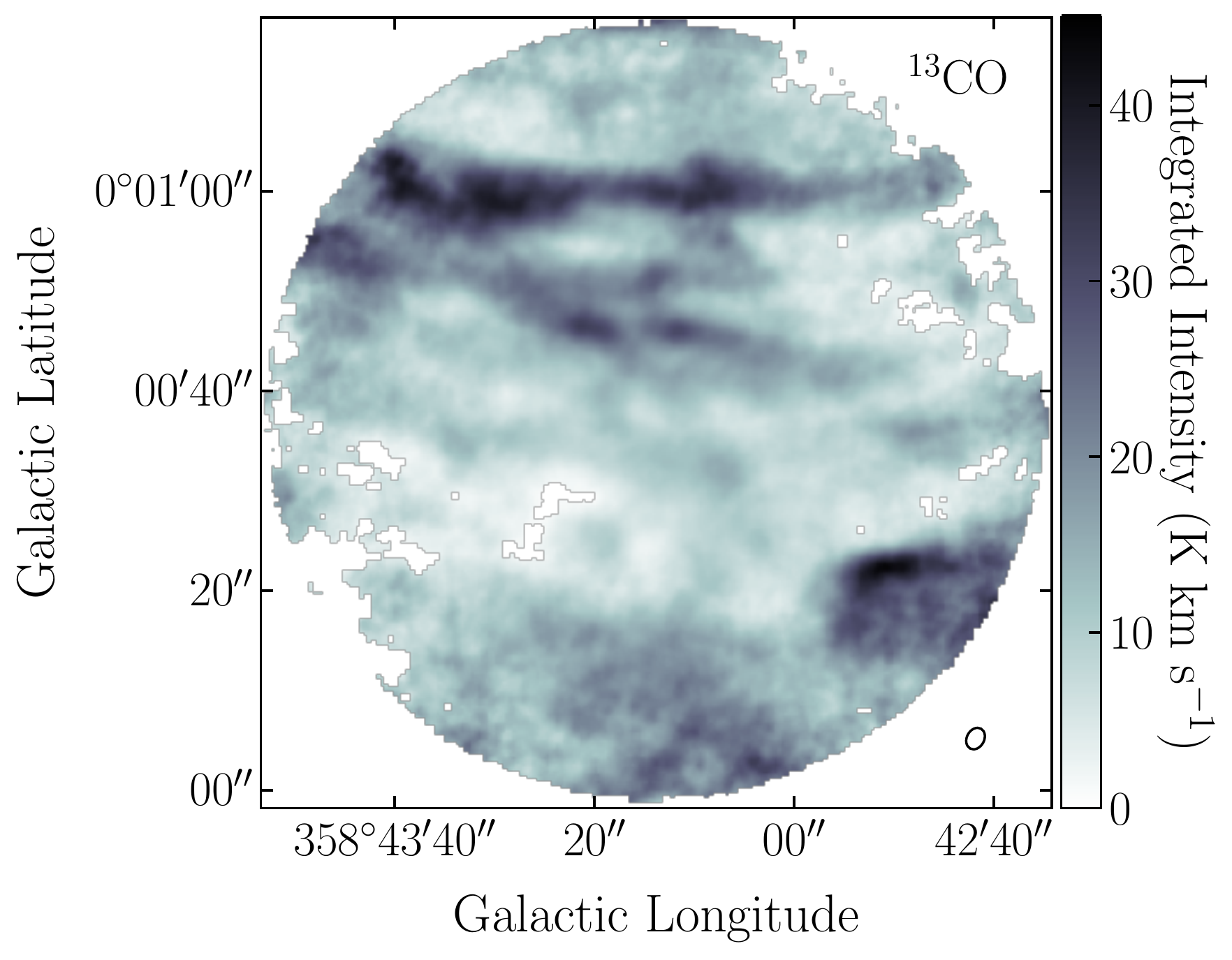}}
\subfigure{
\plotone{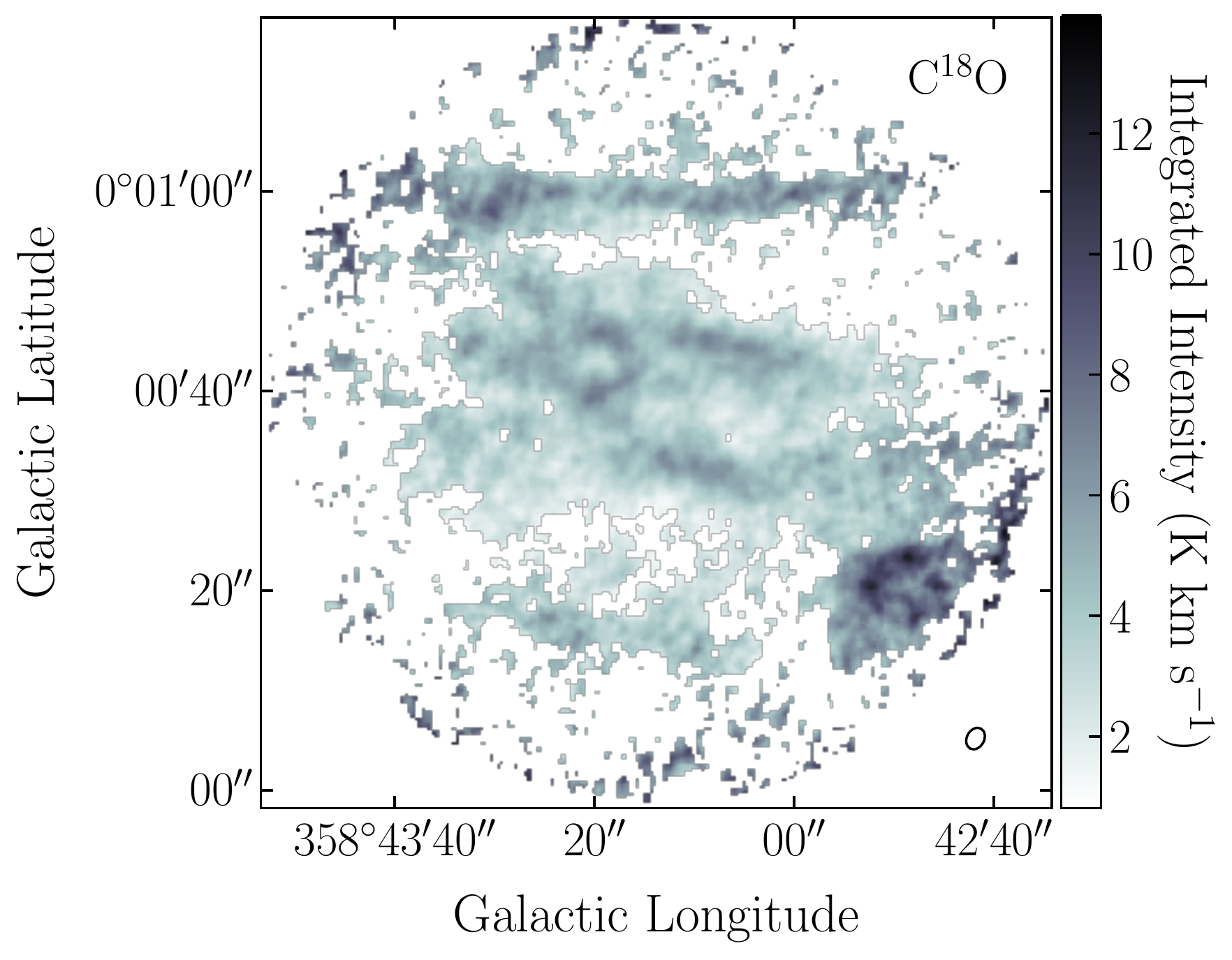}}
\caption{Spatially masked moment 0 maps generated from data in the velocity range of -197 km s$^{-1}$ to -231 km s$^{-1}$ for $^{12}$CO (\textit{top}), $^{13}$CO (\textit{middle}), and C$^{18}$O (\textit{bottom}). In the $^{12}$CO image, the orange line represents the filament 1 \textit{p-v} slice path,  while the blue line represents the filament 2 \textit{p-v} slice path. For each image, the beam is indicated in the lower right corner.}
\label{fig:mom0}
\end{figure}

\begin{figure}[htb!] 
\epsscale{1.05}
\begin{centering}
\subfigure{
\plotone{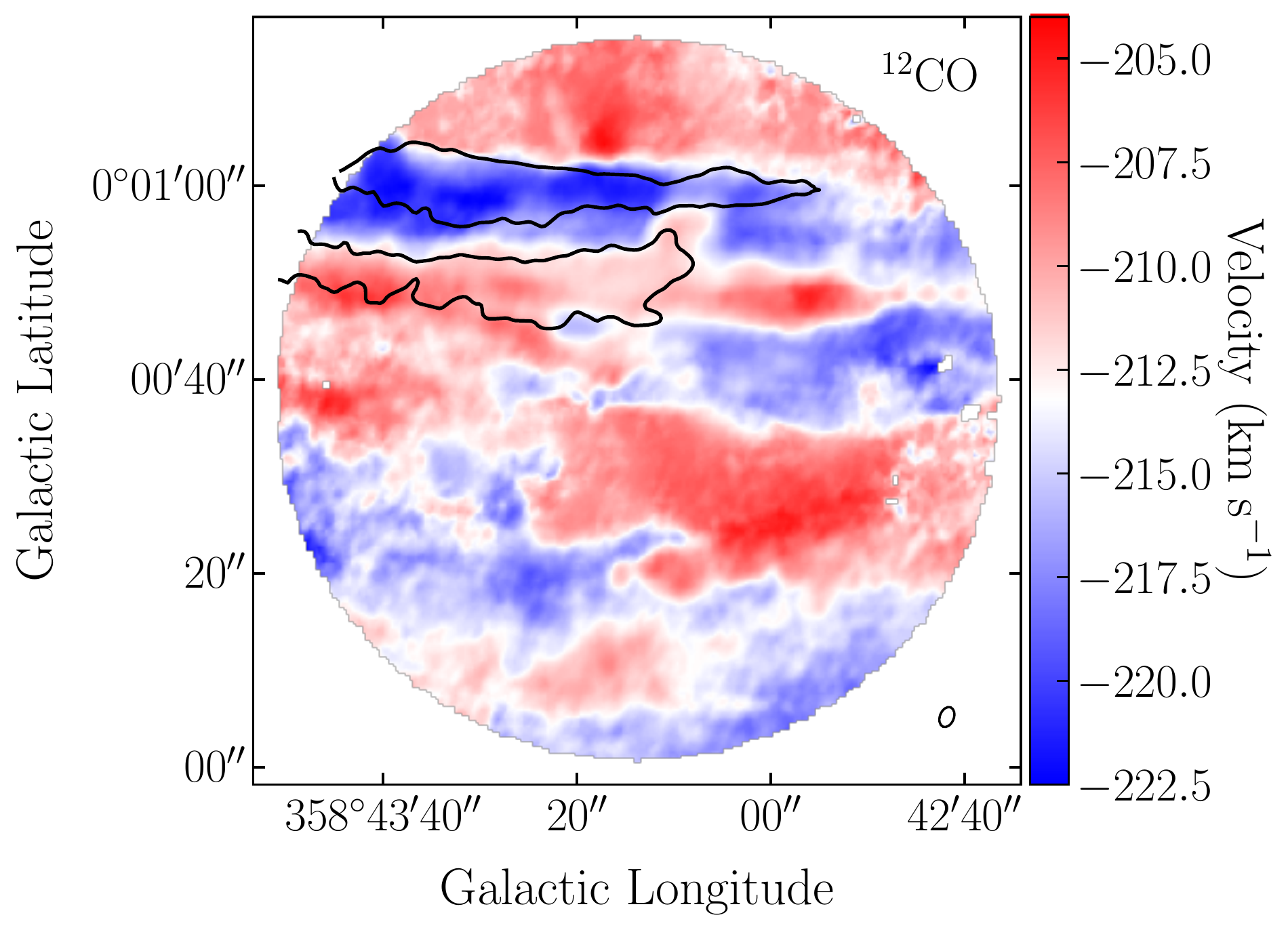}}
\subfigure{
\plotone{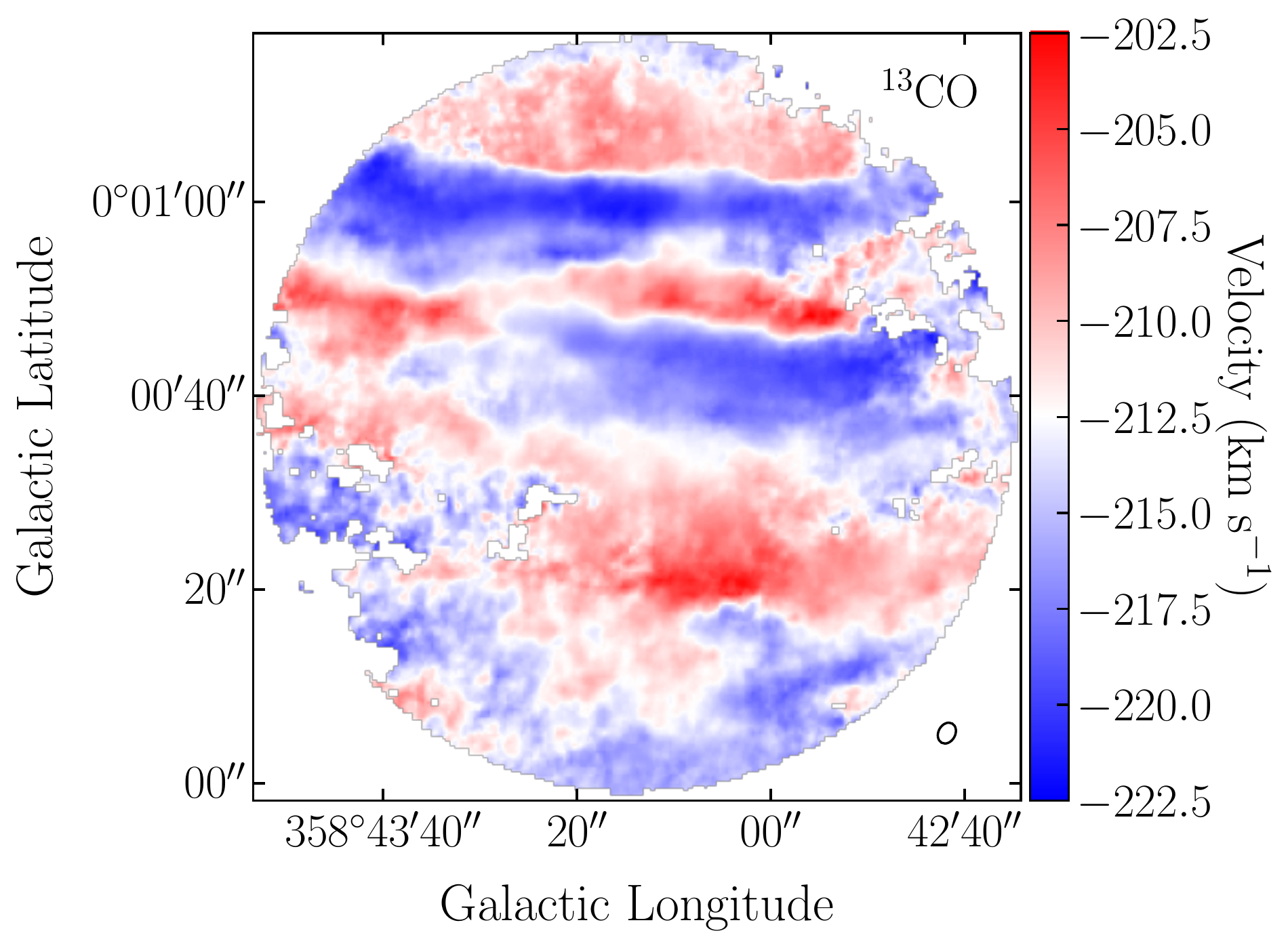}}
\subfigure{
\plotone{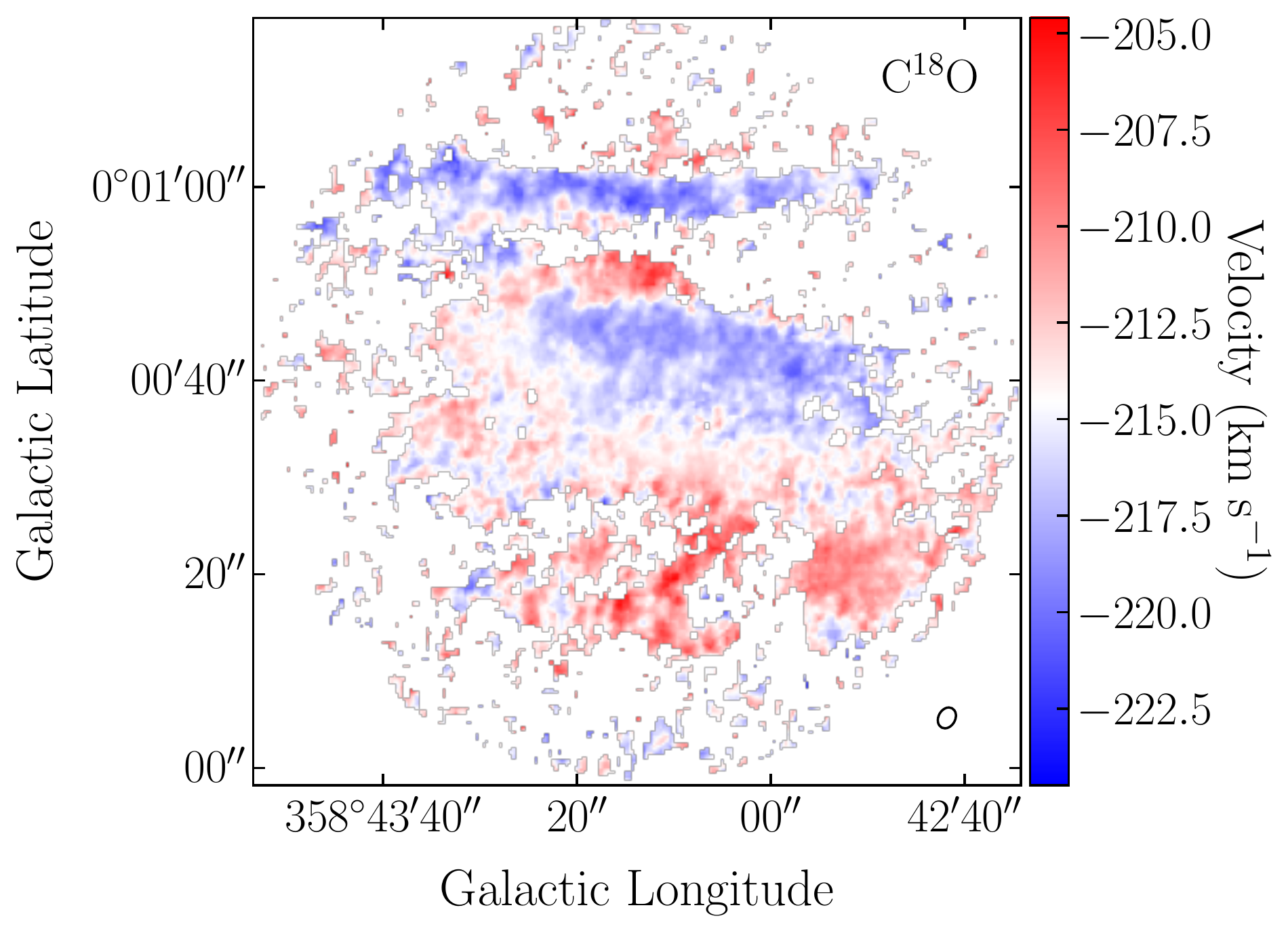}}
\caption{Spatially masked moment 1 maps generated from data subcubes in the velocity range of -197 km s$^{-1}$ to -231 km s$^{-1}$ for $^{12}$CO, $^{13}$CO and C$^{18}$O, respectively. In the $^{12}$CO image, the black contours indicate the filament 1 and filament 2 regions. For each image, the beam is indicated in the lower right corner.}
\label{fig:mom1}
\end{centering}
\end{figure}

To further explore the general physical and kinematic properties of the filaments, we performed moment analyses for each spectral line cube in the velocity range of -197 km s$^{-1}$ to -231 km s$^{-1}$ (see Figures \ref{fig:mom0} and \ref{fig:mom1}). We present these results with the understanding that although moment analysis provides a standard way of parameterizing the spatial and kinematic properties of the gas observed in the 3D data cubes, it does have limitations with regard to interpretation as it condenses three dimensions of information into a two dimensional representation and it can obfuscate complex spectral line structure \citep{Henshaw_2016}.

Before performing any moment calculations, we masked each data cube to remove the low signal-to-noise regions in our images. We first generated a two-dimensional mask which only included pixels with a maximum intensity  along the spectral axis that was three times greater than the standard deviation noise level determined from a line-free region of the spectrum at each pixel. Additionally, we applied a three-dimensional mask to each data set which excluded regions where the signal was less than the 1$\sigma$ noise level. 

With both masks applied to the corresponding data set, we proceeded to calculate the zeroth order and first order moment maps for each isotopologue. These calculations were performed using the Astropy \verb|SpectralCube| \citep{adam_ginsburg_2019_3558614} Python package. This package performs moment analysis using the standard methods, which we briefly describe in this section.

% Moment 0 Maps:

We calculated the zeroth order moment for each isotopologue by integrating the intensity along the spectral axis of the data cube at each pixel. Moment 0 maps for all three spectral lines are shown in Figure \ref{fig:mom0}. We see that in the Galactic latitude range spanning \textit{b} = 0\degree00\arcmin40\arcsec - 0\degree01\arcmin00\arcsec there are linear filamentary molecular gas structures which appear to align closely with the orientation of the Galactic plane. Although these filaments appear to exist in the same regions for each isotopologue, it can be seen that the spatial extent of the structures seen in $^{13}$CO and C$^{18}$O emission are  visually more similar to each other than to the $^{12}$CO emission. This is likely due to the $^{12}$CO spectral line having a higher optical depth, a property that we discuss in more detail in section \ref{subsection:opacity}. We see from the $^{13}$CO and C$^{18}$O data that the filament 1 region (located at \textit{b} $\sim$ 0\degree01\arcmin00\arcsec) seems to contain the largest amount of emission in this velocity range, although this is less apparent when we look at the $^{12}$CO moment 0 map. The emission from the filament 2 region (located around \textit{b} $\sim$ 0\degree00\arcmin50\arcsec) is overall less bright than the filament 1 region. 

% Moment 1 Maps:

We also calculated the intensity-weighted velocity of the spectral line, also known as the first order moment. These moment 1 maps are shown in Figure \ref{fig:mom1} where we see that most of the material located in filament 1 is observed at velocities between -220 km s$^{-1}$ to -223 km s$^{-1}$. On the other hand, most of the emission observed in the filament 2 region spans velocities between -202 to -205 km s$^{-1}$.  These trends are observed in all three moment 1 maps. 

\subsection{Filament Properties}
\label{subsection:fil_properties}

The next phase of analysis focuses on quantifying the specific physical and kinematic characteristics of filaments 1 and 2. In Section \ref{subsubsection:phys_char}, we present radial brightness temperature profiles of each filament created using the \verb|RadFil| package, and estimate their length, width, and orientation with respect to the Galactic plane. In Section \ref{subsubsection:fil_velocity_analysis}, we investigate the velocity structure of each filament. We take position-velocity (\textit{p-v}) slices along both filaments and identify four distinct velocity components. In addition to this, we create velocity profiles for each component in \textit{p-v} space using \verb|RadFil|. We then estimate the velocity gradients and FWHM widths from these results. To round out our investigation of the kinematic properties, we also present averaged spectra for all three spectral lines taken from the filamentary region.

For our analysis, we consider how optical depth in our data might affect the accurate determination of filament properties. In the Galactic center, the spectral line ratio $^{12}$CO/$^{13}$CO $\sim$ 20, while the ratio $^{12}$CO/C$^{18}$O $\sim$ 250 \citep{wilson_rood_1994}. Since C$^{18}$O (1-0) emission is the least abundant among the three isotopologues, it is the least likely to be optically thick. In light of this, we use the C$^{18}$O data to generate our brightness temperature profiles and our velocity profiles in an effort to make estimates that are as accurate as possible. We discuss the effects of optical depth on our calculations in greater detail in Section \ref{subsection:opacity}.

\subsubsection{Physical characteristics}
\label{subsubsection:phys_char}

% brightness temperature Profiles:
\begin{figure}[htb!] 
\epsscale{1.0}
\begin{centering}
\subfigure{
\plotone{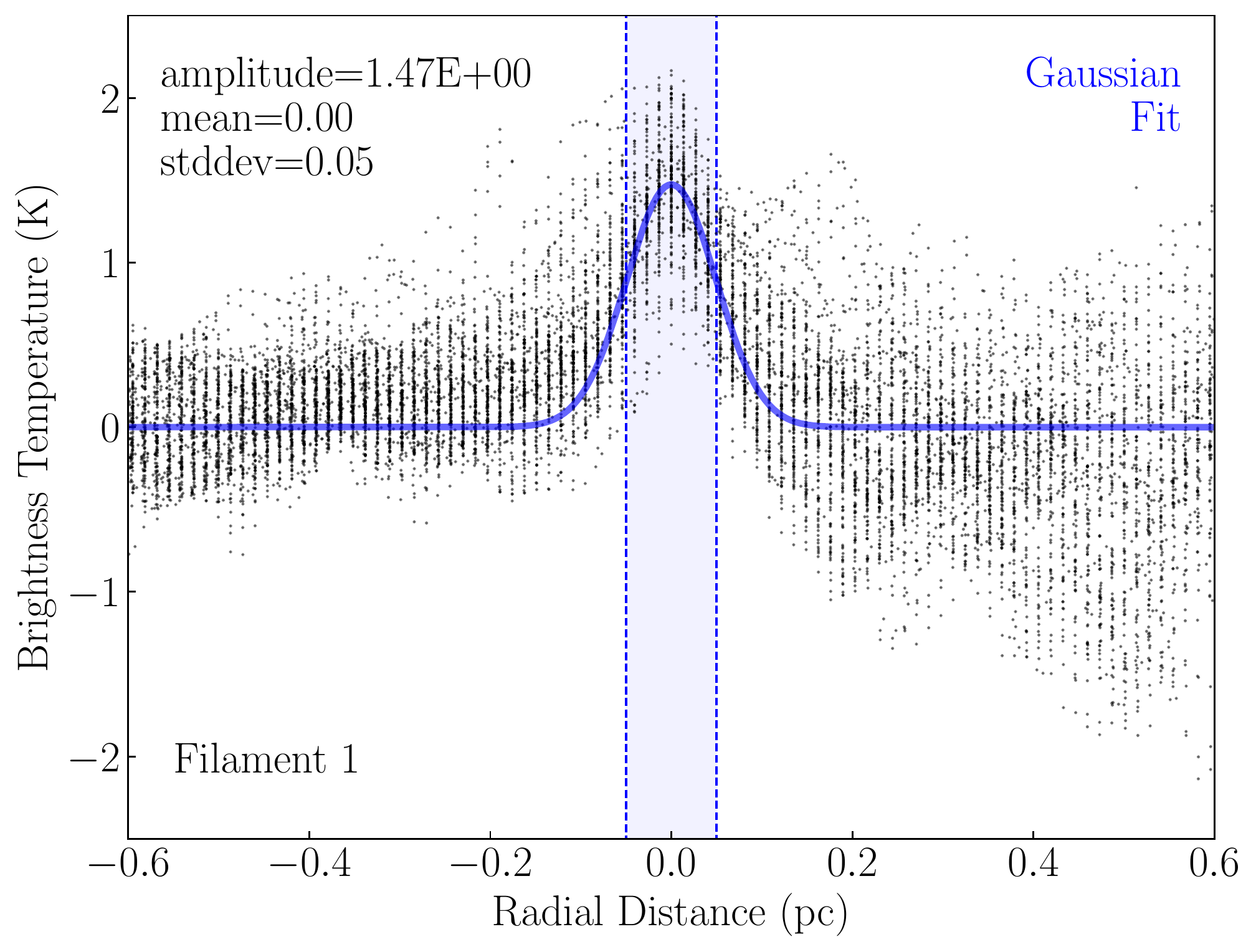}}
\subfigure{
\plotone{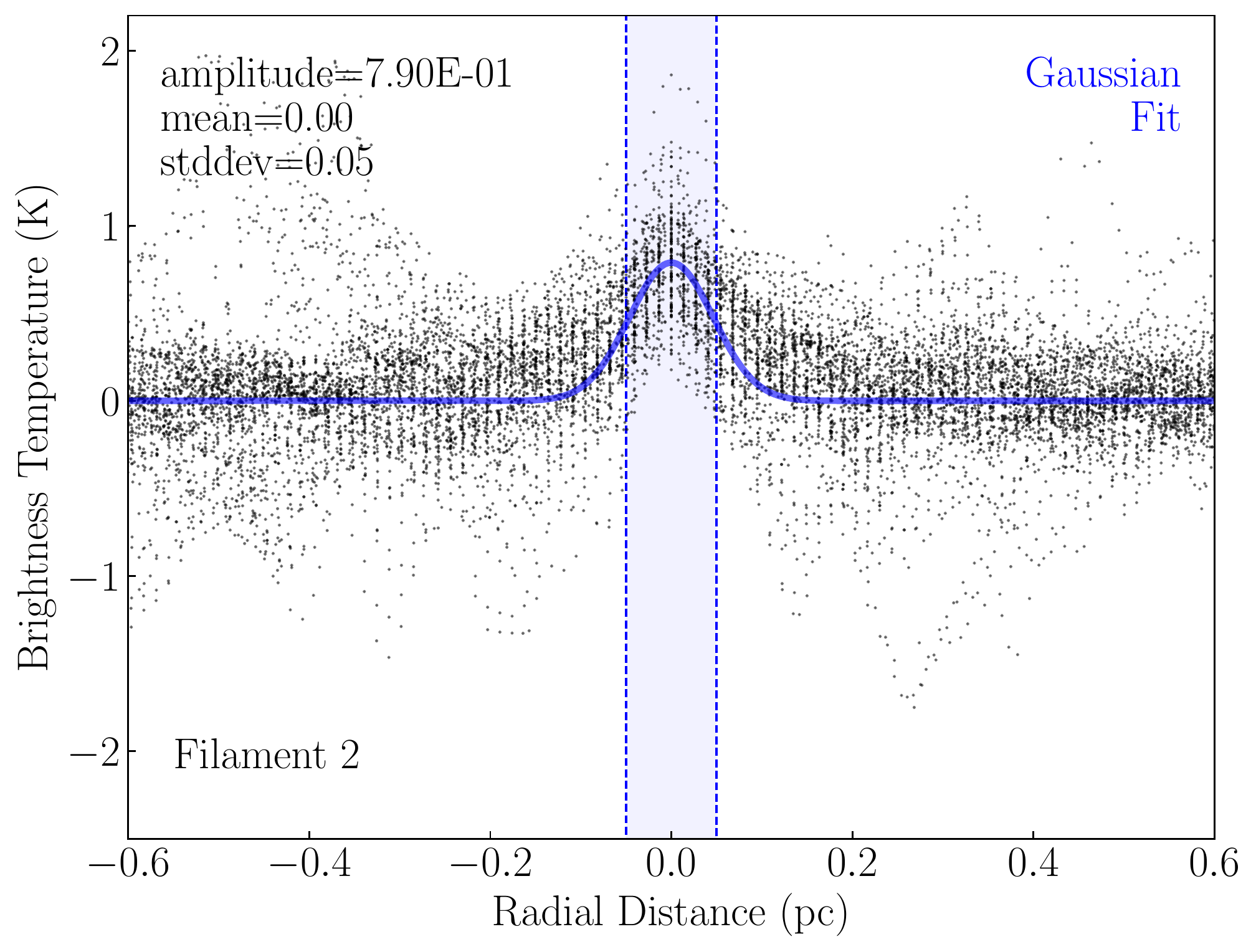}}
\caption{(\textit{Top}) The brightness temperature profile for filament 1 taken from the -223.0 km s$^{-1}$ C$^{18}$O channel image. (\textit{Bottom}) The brightness temperature profile for filament 2 from the -203 km s$^{-1}$ C$^{18}$O channel image. The blue line plotted in both figures is the Gaussian fit for each filament profile made using a fitting distance of 0.05 pc from the center of the filament. The radial distance is the distance from the center of the filament.}
\label{fig:rad_prof}
\end{centering}
\end{figure}

% Table of physical properties:
\begin{table*}[htb!] 
\centering
\caption{Physical properties for filaments 1 and 2.}
\begin{tabular*}{0.70\textwidth}{c c c c c}
\hline\hline
Filament & Spine Length (pc) & FWHM Width (pc) & Aspect Ratio & Position Angle (\degree)\footnote{The absolute 2D projected position angle between the filament and the Galactic midplane.}\\ 
\hline
1 & 2.2 & 0.12$\pm{0.02}$ & 4.5:1 & 1.8 \\ % C18O values
2 & 1.9 & 0.11$\pm{0.01}$ & 5.4:1 & 1.2 \\ % C18O values
\hline
\end{tabular*}
\label{table:fil_prop}
\end{table*}

We investigated the physical properties of the Sgr E filaments primarily with the use of the Python packages \verb|RadFil| \citep{Zucker_2018}\footnote{https://github.com/catherinezucker/radfil} and \verb|FilFinder| \citep{2015MNRAS.452.3435K}\footnote{https://github.com/e-koch/FilFinder}. We did this by applying a boolean mask of the filament 1 and filament 2 regions (see Section \ref{subsubsection:fil_ID}) onto the -223 km s$^{-1}$ and -203 km s$^{-1}$ C$^{18}$O channel images, respectively (Figure \ref{fig:chan18_fil}). We then created a one-pixel wide filament spine for each filament in the masked images through the use of \verb|RadFil|, which calls upon the \verb|FilFinder| package. \verb|FilFinder|  produces this spine object by performing medial axis skeletonization on the filament region, a process that finds the set of all pixels within the given mask that has more than
one closest point on the shape’s boundary. This makes it so that each pixel contained within the spine object is at a maximum distance it can be from the edge of the filamentary region. A more detailed explanation of the medial axis skeletonization, or medial axis transformation, process can be found in \cite{2015MNRAS.452.3435K}.

Once the spines were produced, we used \verb|RadFil| to calculate the length of the each filament (see Table \ref{table:fil_prop}). From here, we calculated a linear fit from the spine object in pixel coordinates using  \verb|np.polyfit|. We then used this line to estimate the angle of the filament from the Galactic plane in degrees, which we report in Table \ref{table:fil_prop}. 

We then used \verb|RadFil| to produce brightness temperature profiles for both filaments, shown in Figure \ref{fig:rad_prof}. To do this, \verb|RadFil| smooths the filament spine and takes evenly-sampled cuts for which a brightness temperature profile perpendicular to the spine is calculated at each site along the filament. A thorough description of this process is given in \cite{Zucker_2018}. We used a sampling interval of 1 pixel along the spine ($\sim$ 0.01 pc). We created the filament 1 brightness temperature profile from the C$^{18}$O -223 km s$^{-1}$ channel image and the filament 2 profile from the C$^{18}$O -203 km s$^{-1}$ channel image. As seen in Figure \ref{fig:rad_prof}, we generated Gaussian fits from the brightness temperature profiles for each filament using \verb|RadFil|. For both filaments, we used a fitting distance of 0.05 pc from the center of the filament, where the fitting distance is the region for which the Gaussian fit is calculated.  We then estimated the filament widths in parsecs using the FWHM values of the Gaussian fits, which are recorded in Table \ref{table:fil_prop}. To calculate the filament aspect ratios, we assume a cylindrical geometry for the filaments and set the cross-sectional area of a cylinder equal to the area under the Gaussian fits for each filament (Figure \ref{fig:rad_prof}). From there, we divide the filament length by the estimated effective filament diameter to acquire the filament aspect ratios.

From these results, we see that both filament 1 and 2 exhibit similar physical dimensions as both have lengths of $\sim$2 pc, FWHM widths of $\sim$ 0.1 pc, and aspect ratios of $\sim$ 5:1. We stress that these are lower limit calculations for the filament lengths and aspect ratios, as the filaments appear to extend out of the field of view. We also note that filament widths of 0.1 pc, similar to our spatial resolution, suggest that these may be unresolved and could be narrower. Both filaments are well aligned with the Galactic plane, with modest position angles of 1.8 (filament 1) and 1.2 degrees (filament 2) and are oriented so that the material at lower Galactic longitudes is closer to the Galactic plane.

\subsubsection{Velocity structure of filaments}
\label{subsubsection:fil_velocity_analysis}

% PV Diagrams for Filament 1
\begin{figure}[htb!] 
\epsscale{1.0}
\begin{centering}
\subfigure{
\plotone{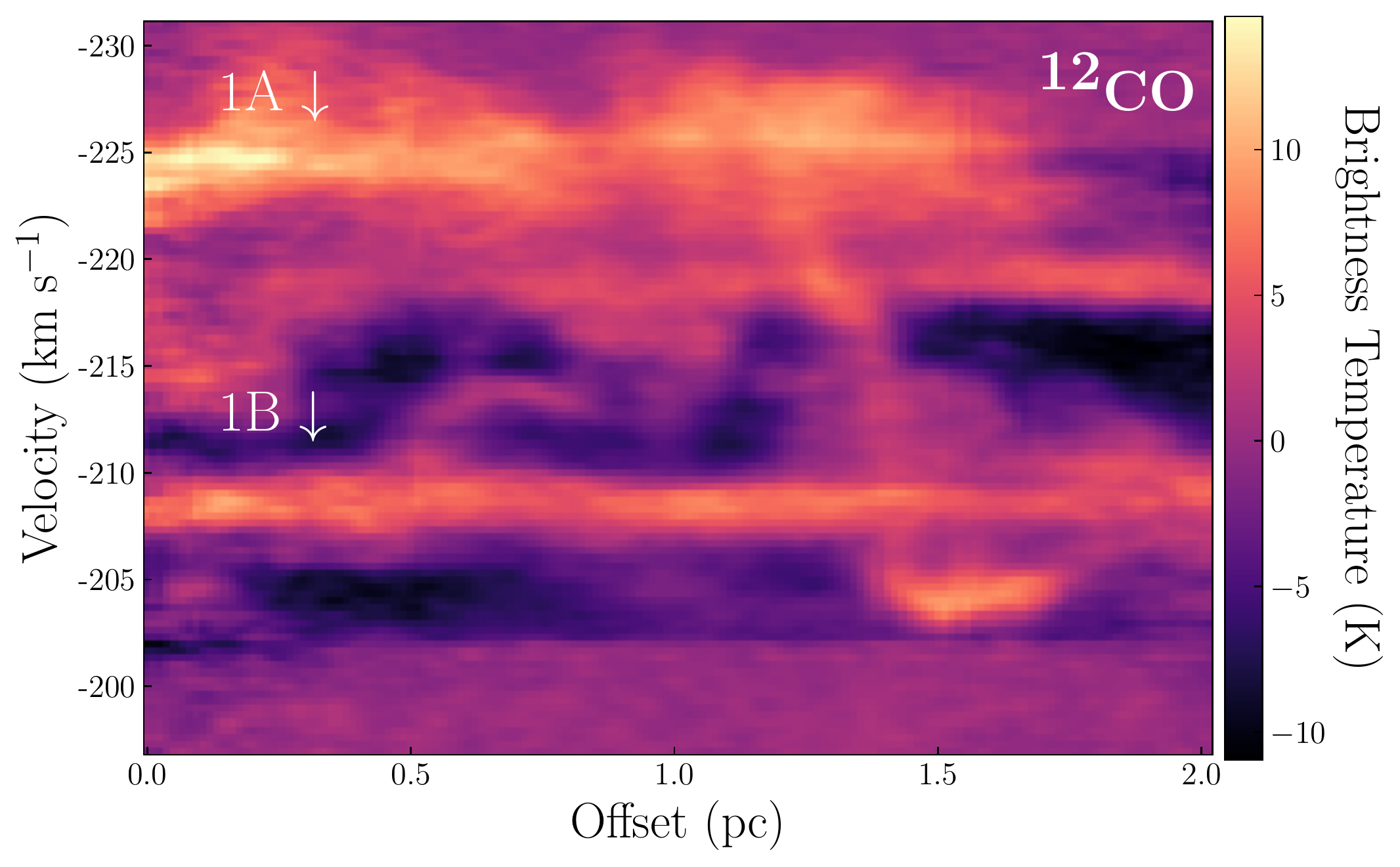}}
\subfigure{
\plotone{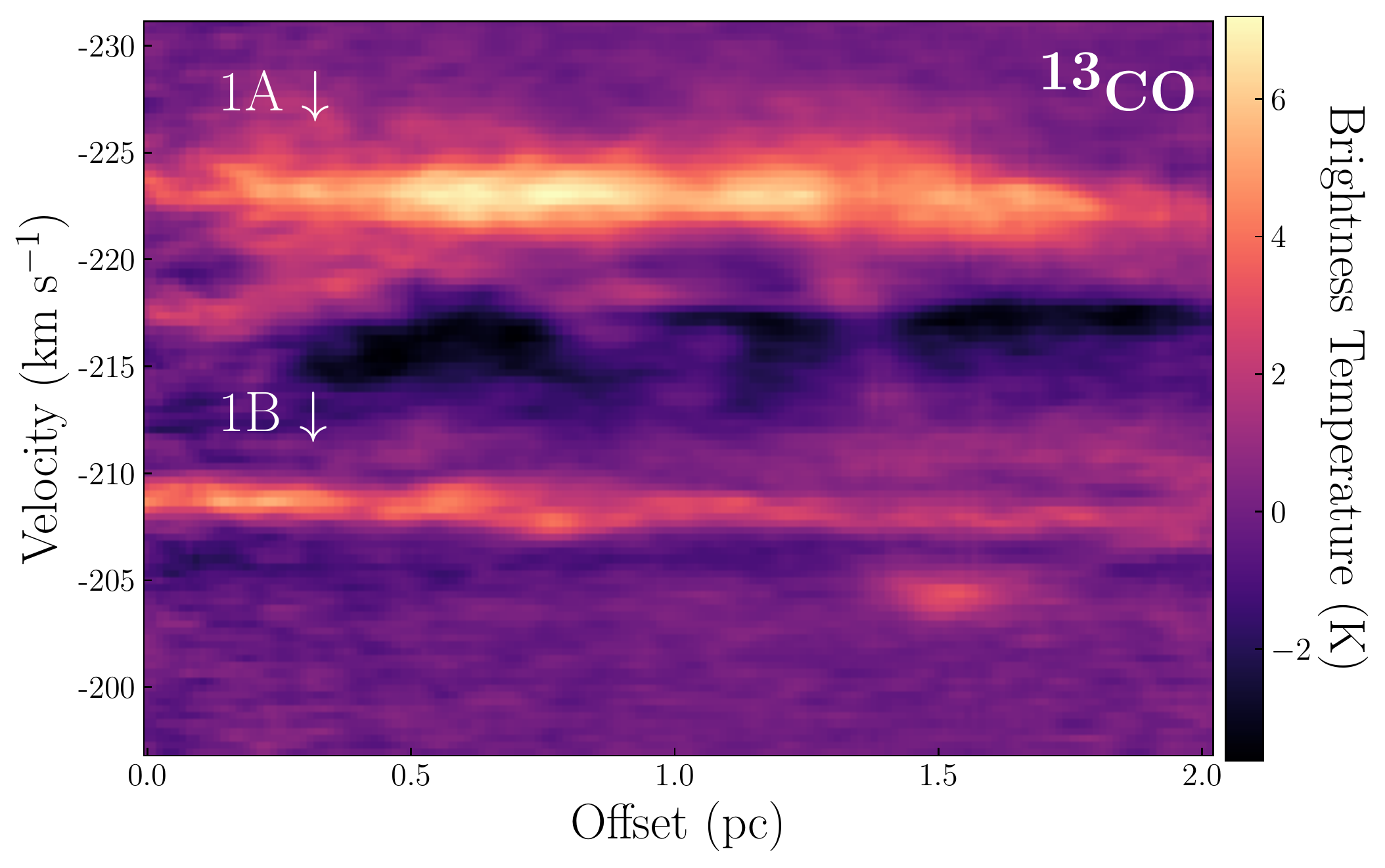}}
\subfigure{
\plotone{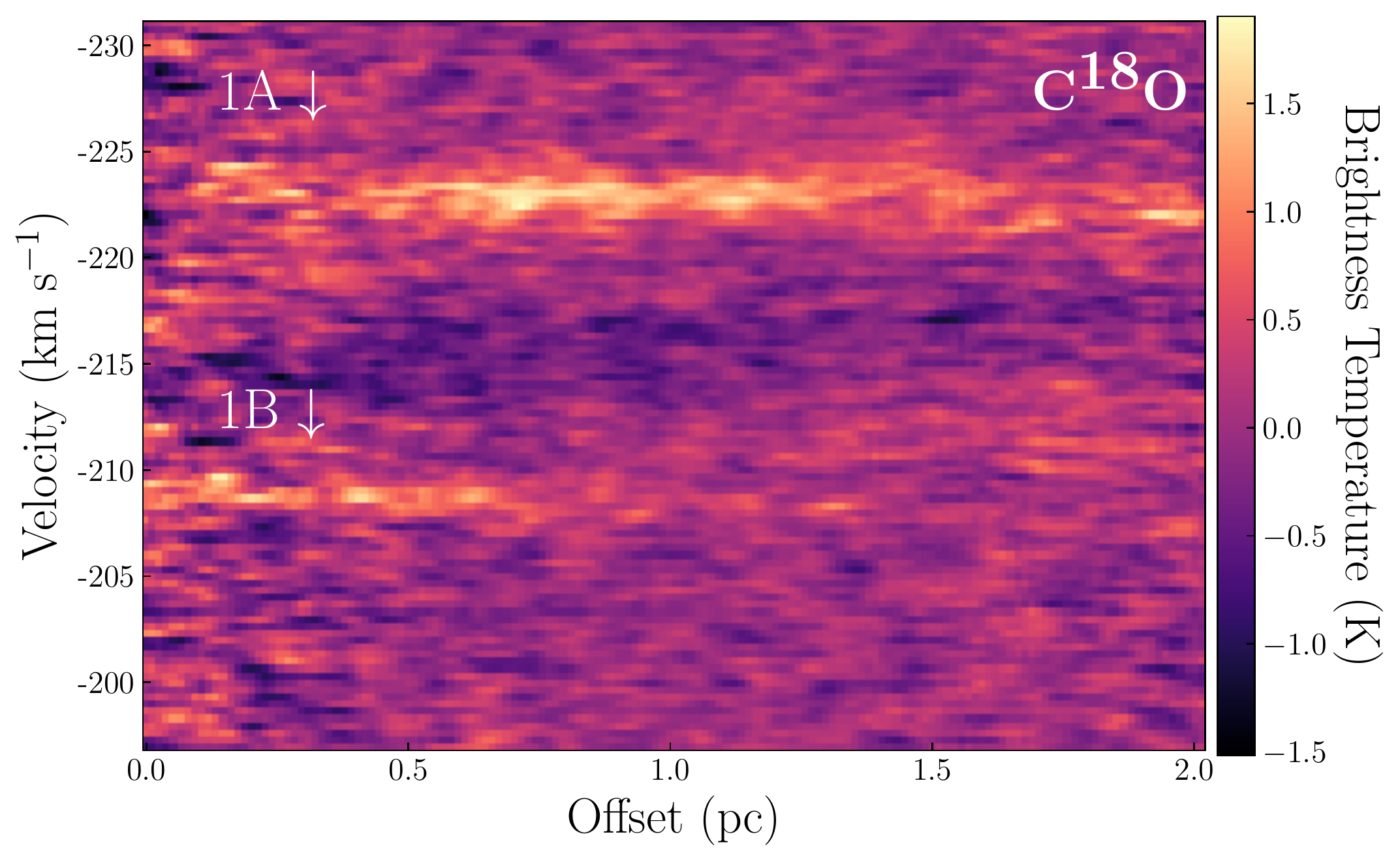}}
\caption{The filament 1 position-velocity diagrams for $^{12}$CO (\textit{top)}, $^{13}$CO (\textit{middle}), and  C$^{18}$O (\textit{bottom)}, respectively. The $p-v$ slice starts on the end of the filament that is closer to the Galactic center, so the offset represents the distance along the $p-v$ slice as we move from left to right along the filament 1 spine (see Figure \ref{fig:mom0}).}
\label{fig:PV_fil_1}
\end{centering}
\end{figure}

% PV Diagrams for Filament 2
\begin{figure}[htb!] 
\epsscale{1.0}
\begin{centering}
\subfigure{
\plotone{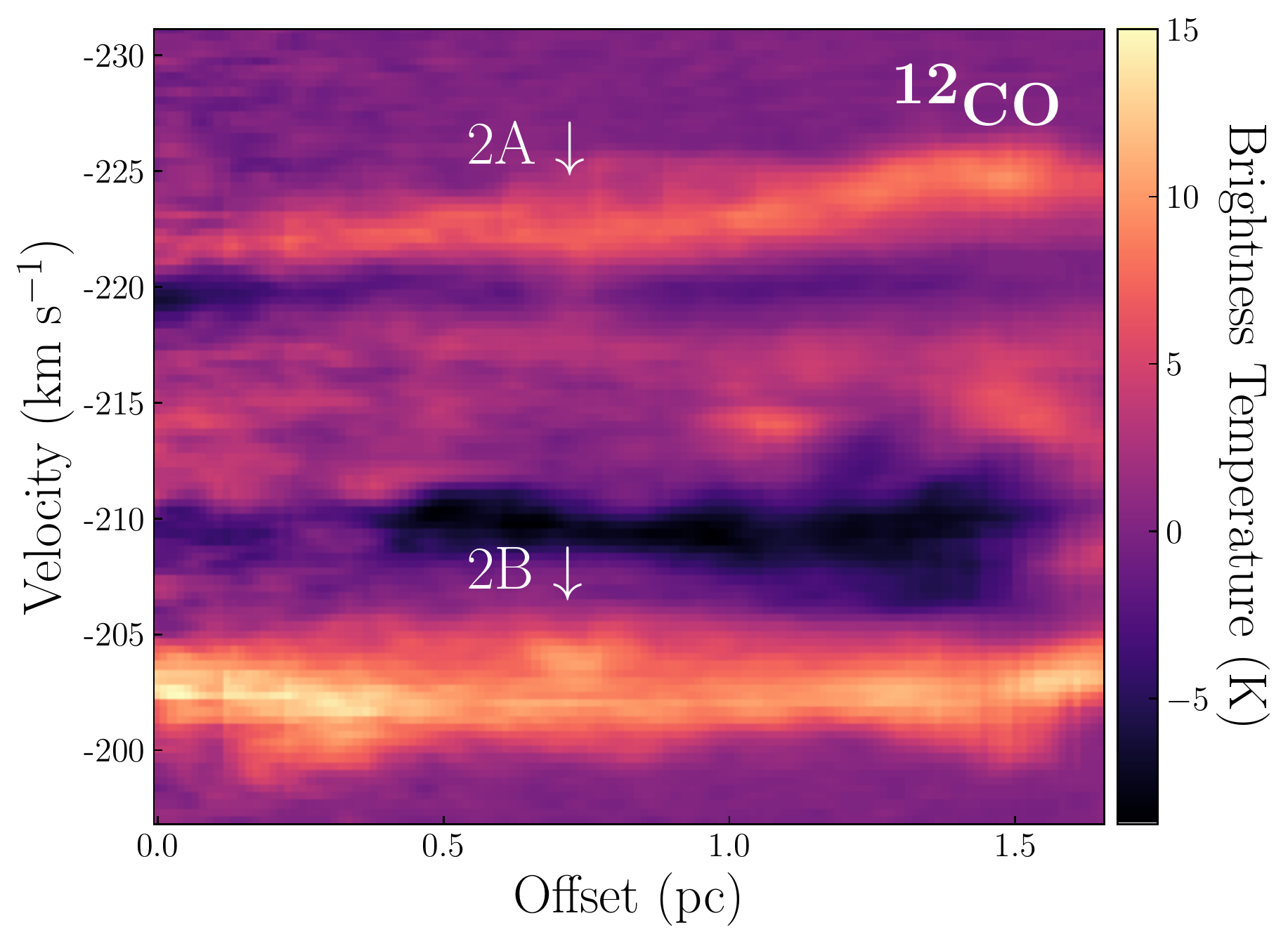}}
\subfigure{
\plotone{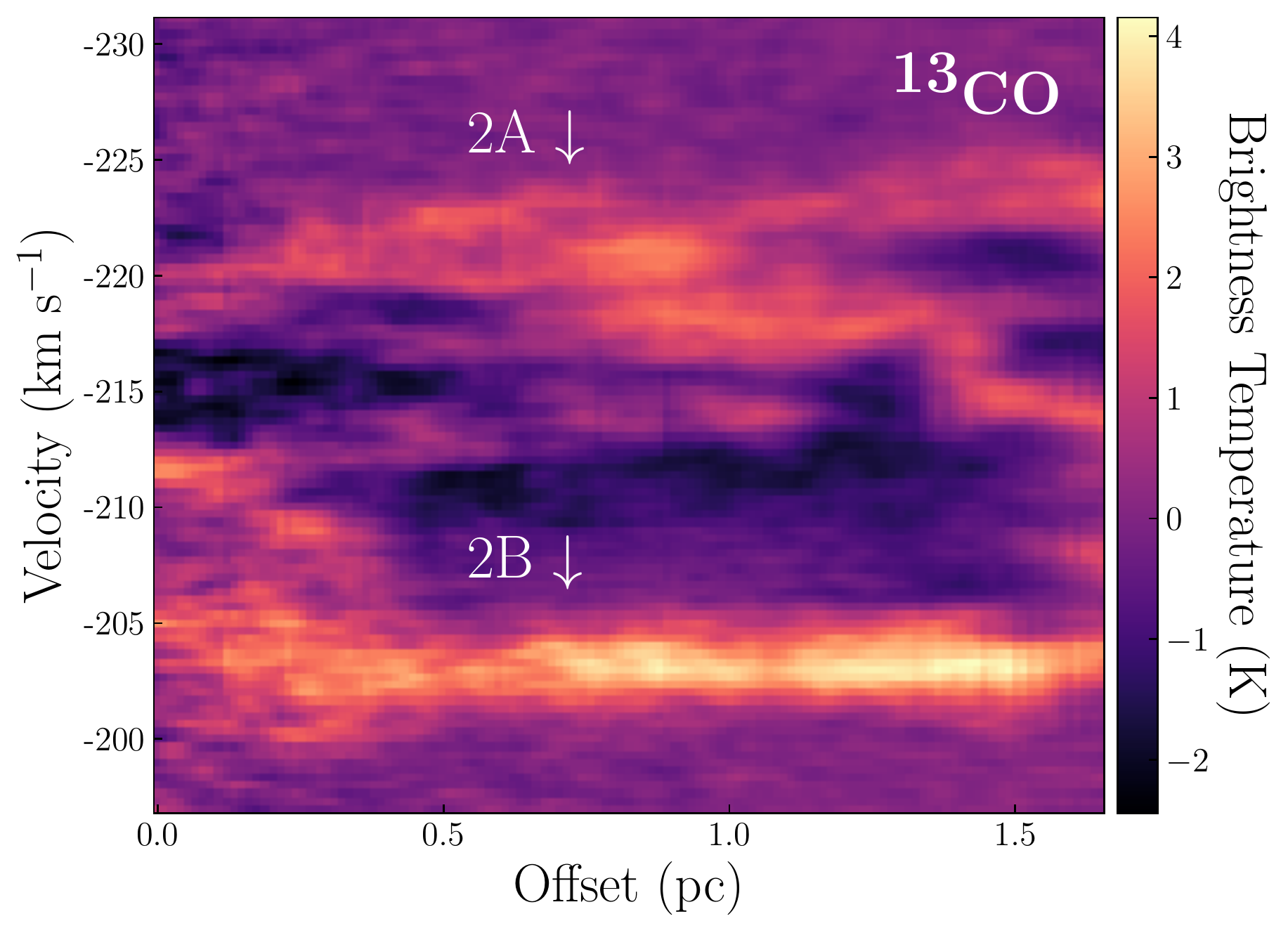}}
\subfigure{
\plotone{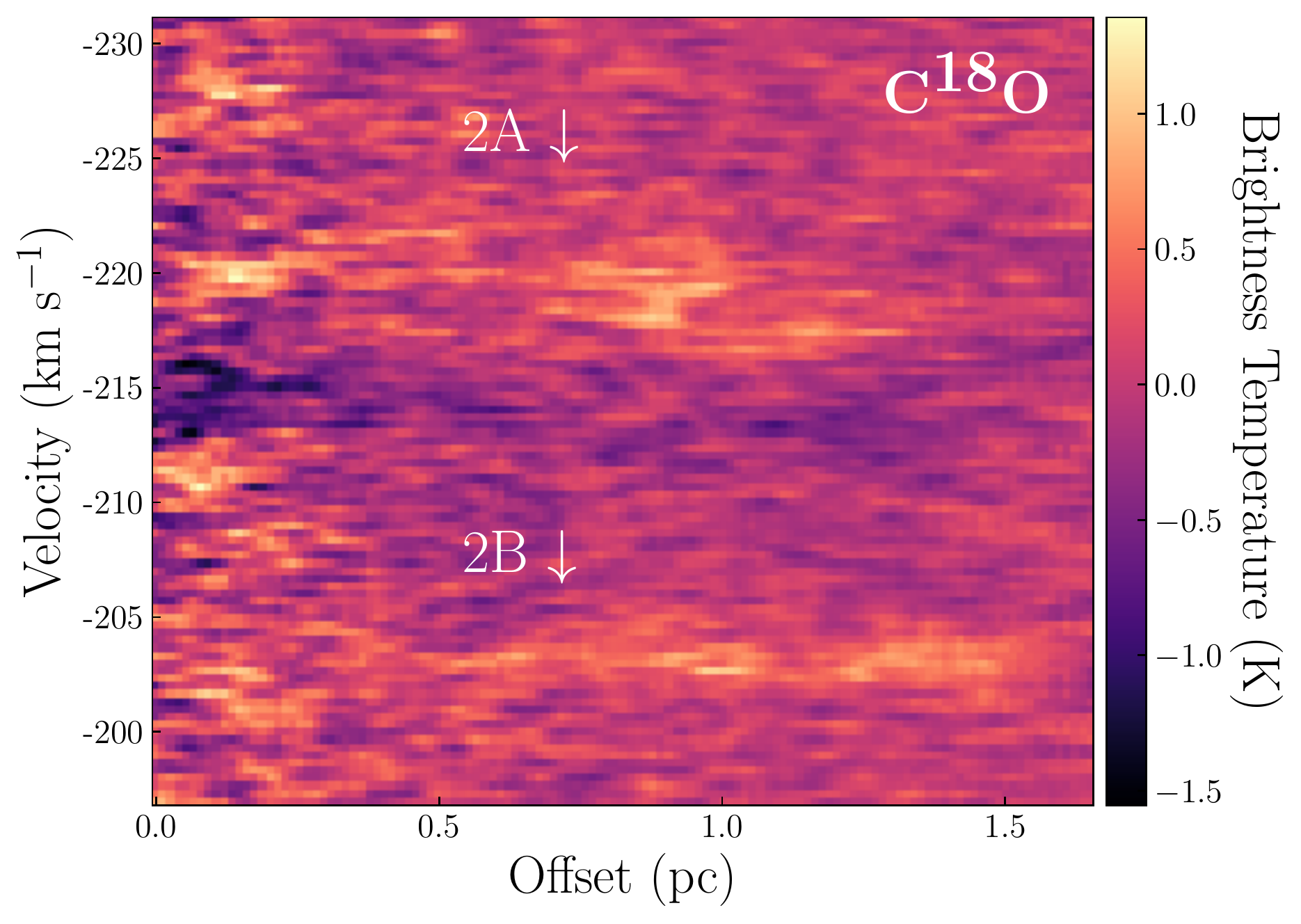}}
\caption{The filament 2 position-velocity diagrams for $^{12}$CO (\textit{top)}, $^{13}$CO (\textit{middle}), and  C$^{18}$O (\textit{bottom)}, respectively. The $p-v$ slice starts on the end of the filament that is closer to the Galactic center, so the offset represents the distance along the $p-v$ slice as we move from left to right along the filament 2 spine (see Figure \ref{fig:mom0}).}
\label{fig:PV_fil_2}
\end{centering}
\end{figure}

We take position-velocity (\textit{p-v}) slices along filaments 1 and 2 for all three data cubes in the range of -197 km s$^{-1}$ to -231 km s$^{-1}$ through the use of the \verb|pvextractor| package \citep{ginsburg_pv_2016}. The \textit{p-v} slices were made along the filament spines calculated using \verb|RadFil|. By convention, the path begins (offset = 0 pc) at higher Galactic longitudes and moves to lower longitudes (left to right in the figures). The paths for the ${p-v}$ slices are shown in Figures \ref{fig:mom0} and \ref{fig:mom1} as orange (filament 1) and blue (filament 2) lines. We calculated the \textit{p-v} slices with a given width of 2\arcsec ($\sim$ 0.09 pc), about the same size as the semimajor axis of our beam.

Although the purpose of the \verb|RadFil| package is to calculate radial profiles in position-position space, as we described in Section \ref{subsubsection:phys_char}, for this project we also used it in position-velocity space to determine certain kinematic properties of the filaments, namely their velocity gradients and velocity widths. We again used the \verb|FilFinder| package to create a `spine' for the brightest region calculated along the aforementioned \textit{p-v} slice. Just as before, we generated a linear fit of the spine for each notable velocity component. We then calculated the velocity gradient of the filament by determining the slope of the resulting best-fit line. We use the median value of the pixel coordinates of the best-fit line to estimate the central velocity for each component.

In the \textit{p-v} diagram sliced along filament 1 (Figure \ref{fig:PV_fil_1}), we observe two distinct velocity components that are seen in all three CO isotopologues. We label these regions as velocity components 1A and 1B. These components are separated by $\sim$ 14 km s$^{-1}$ along the length of the filament. We take velocity component 1A to be the component that is associated with filament 1 as it is centered around -223 km s$^{-1}$ and is the brightest of the two in each diagram, especially with regard to the $^{13}$CO and C$^{18}$O emission.  In comparison, the 1B component centered at -209 km s$^{-1}$ is less bright and has a narrower velocity width than the 1A component. From inspection of the data cubes, we believe that velocity component 1B is associated with local non-filamentary CO emission. For $^{12}$CO and $^{13}$CO in Figure \ref{fig:PV_fil_1}, there is a ``bridge'' of emission located at an offset of $\sim$1.0 pc, a feature that could indicate an interaction between filament 1 and the gas associated with velocity component 1B. We note that components 1A and 1B are both present along the full length of the filament and have velocity gradients of 0.35 km~s$^{-1}$~pc$^{-1}$ and 0.57 km~s$^{-1}$~pc$^{-1}$, respectively.  

Similarly, we see in Figure \ref{fig:PV_fil_2} that there are two prominent velocity components along filament 2, which we label 2A and 2B. In the $^{13}$CO and C$^{18}$O \textit{p-v} diagrams, we see that component 2A and 2B have central velocities of -219 km s$^{-1}$ and -203 km s$^{-1}$, respectively. These velocity components are separated by $\sim$ 16 km s$^{-1}$ along the length of filament 2.  Velocity component 2A has a measured velocity gradient of 5.2 km~s$^{-1}$~pc$^{-1}$, while component 2B exhibits a small velocity gradient of 0.0017 km~s$^{-1}$~pc$^{-1}$. In the $^{12}$CO and $^{13}$CO figures, it is clear that component 2B is much brighter than component 2A, however this is more ambiguous in the C$^{18}$O data, which only has a faint signal for both components. Based on its central velocity of -203 km s$^{-1}$, we believe that velocity component 2B is definitely associated with filament 2. The central velocity and velocity gradient of velocity component 2A, leads us to interpret it as the emission associated with the other filament-like feature observed in panels d and e of Figure \ref{fig:mini_mom0}.

Velocity components 1A, 1B, and 2B, the velocity of the gas seems to slowly decrease as we move towards lower Galactic longitudes away from the Galactic center, behavior that is consistent with the large-scale trend along the far dust lane. In Figure \ref{fig:PV_fil_2} it is important to note the apparent absorption at -220 km s$^{-1}$ in the $^{12}$CO \textit{p-v} diagram. This is a feature that is not present in the \textit{p-v} diagrams for $^{13}$CO and C$^{18}$O, and it may be the result of $^{12}$CO emission being optically thick towards this region. 

In a process analogous to the creation of the brightness temperature profile, we calculated a velocity profile along each velocity component we identified. We generated a Gaussian fit to the velocity profiles and recorded the FWHM linewidth. All of these analysis steps were performed for each velocity component using the C$^{18}$O \textit{p-v} diagrams (Figures \ref{fig:PV_fil_1} and \ref{fig:PV_fil_2}). We used the higher signal-to-noise $^{13}$CO \textit{p-v} data to mask the C$^{18}$O data. We used masking thresholds of 2.8 K, 1.8 K, 1.0 K, and 2.1 K for velocity components 1A, 1B, 2A, and 2B, respectively.

\begin{figure*}[htb!] 
%\epsscale{0.75}
\begin{centering}
\subfigure{
\includegraphics[width=0.48\textwidth]{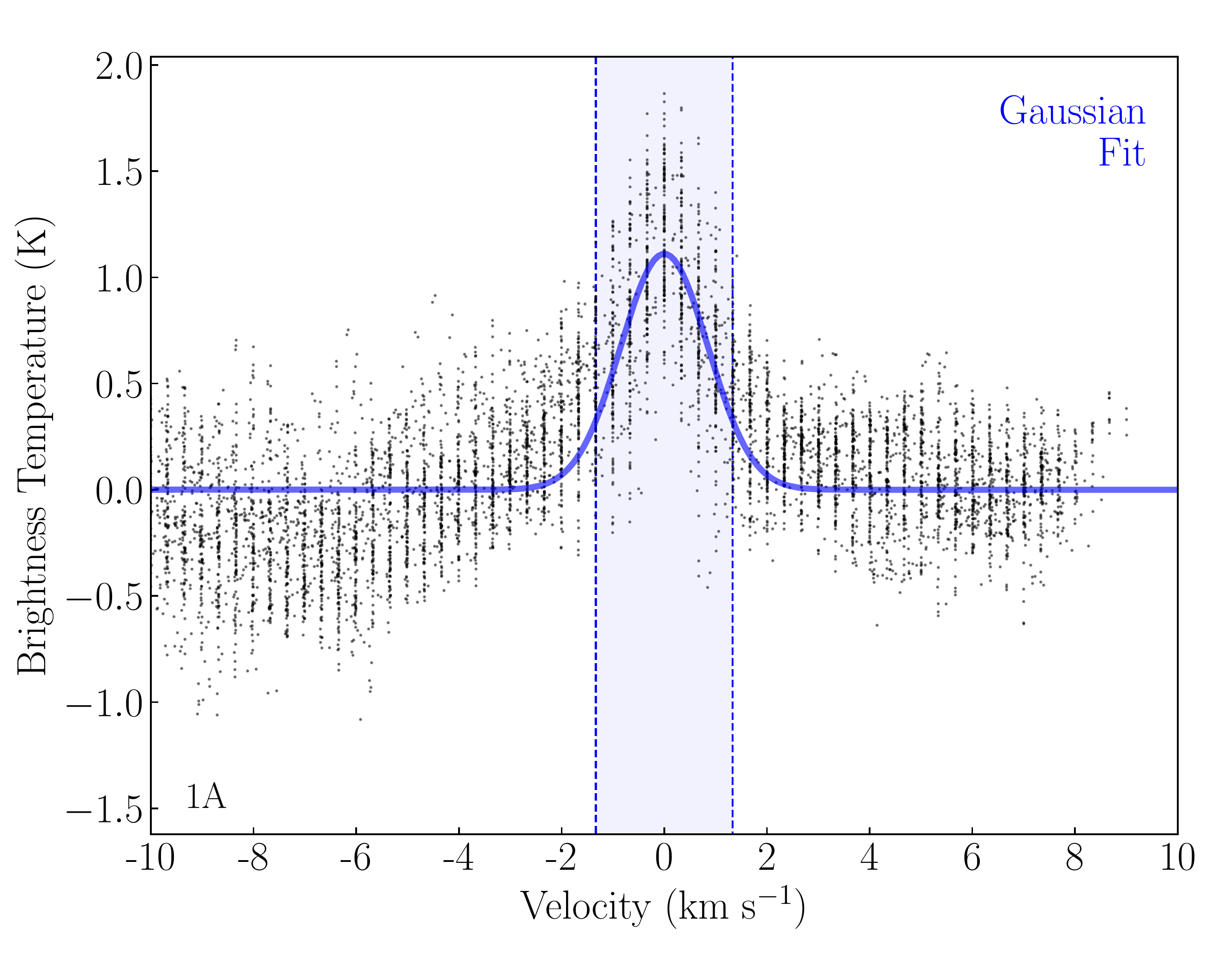}}
\subfigure{
\includegraphics[width=0.48\textwidth]{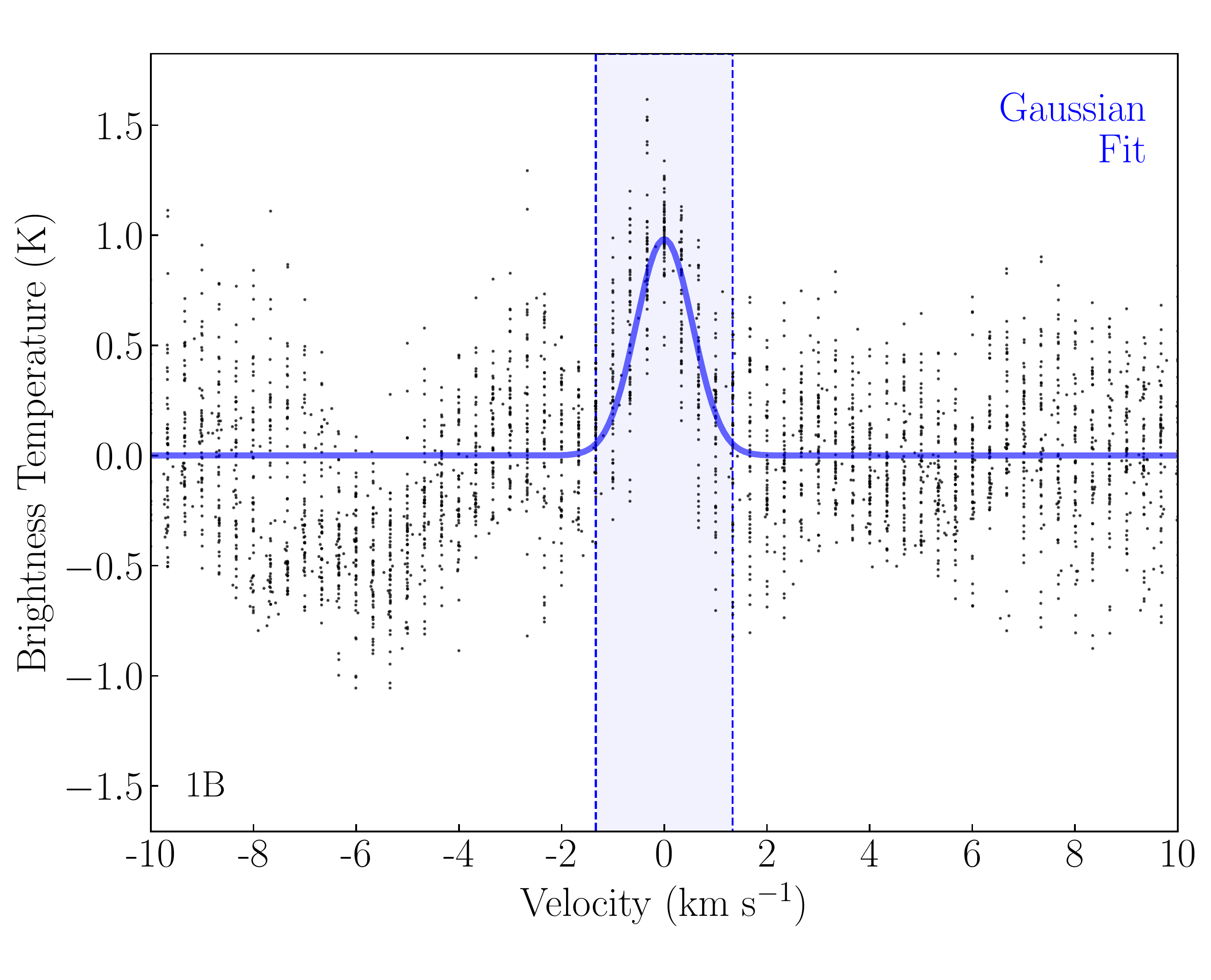}}
\subfigure{
\includegraphics[width=0.48\textwidth]{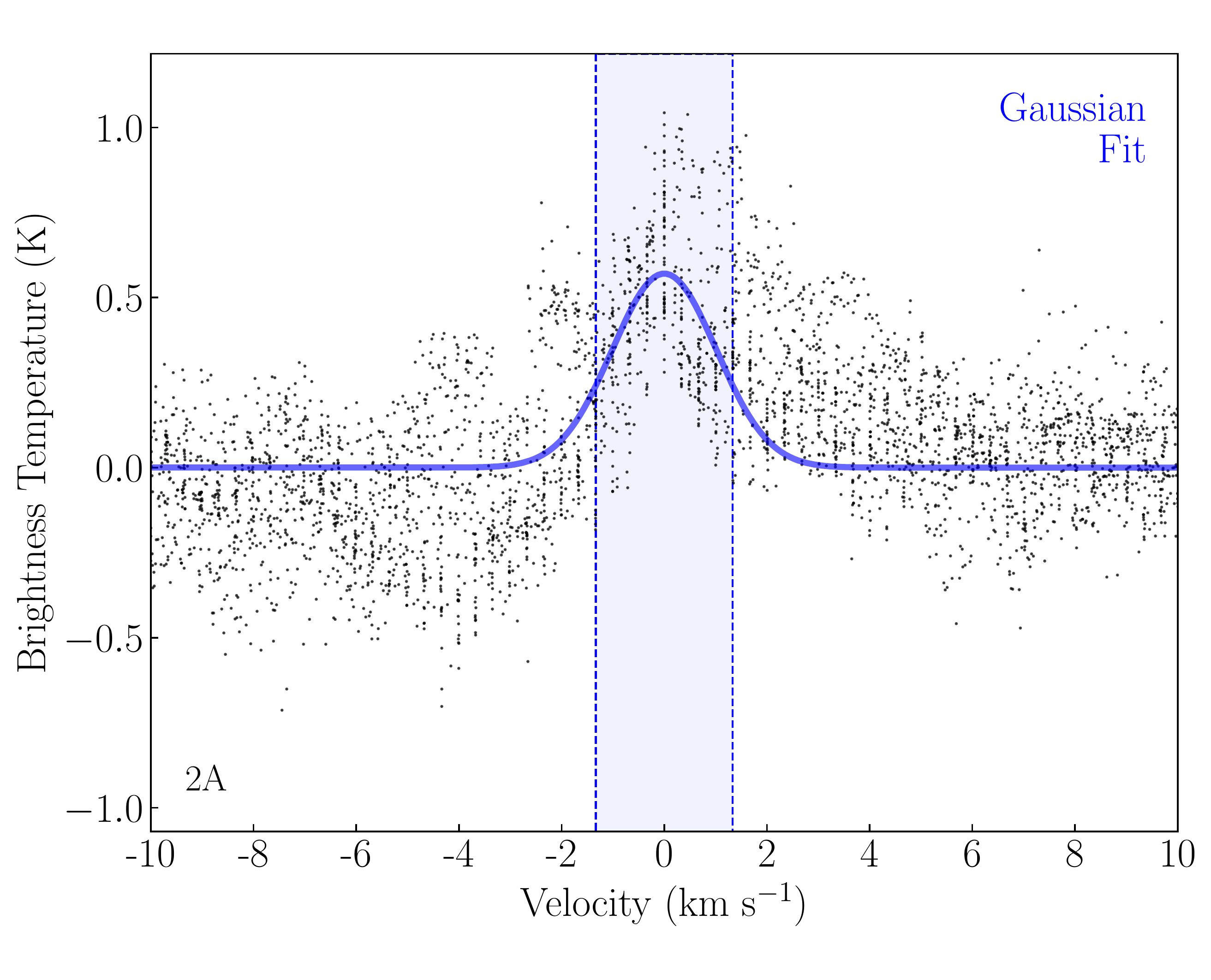}}
\subfigure{
\includegraphics[width=0.48\textwidth]{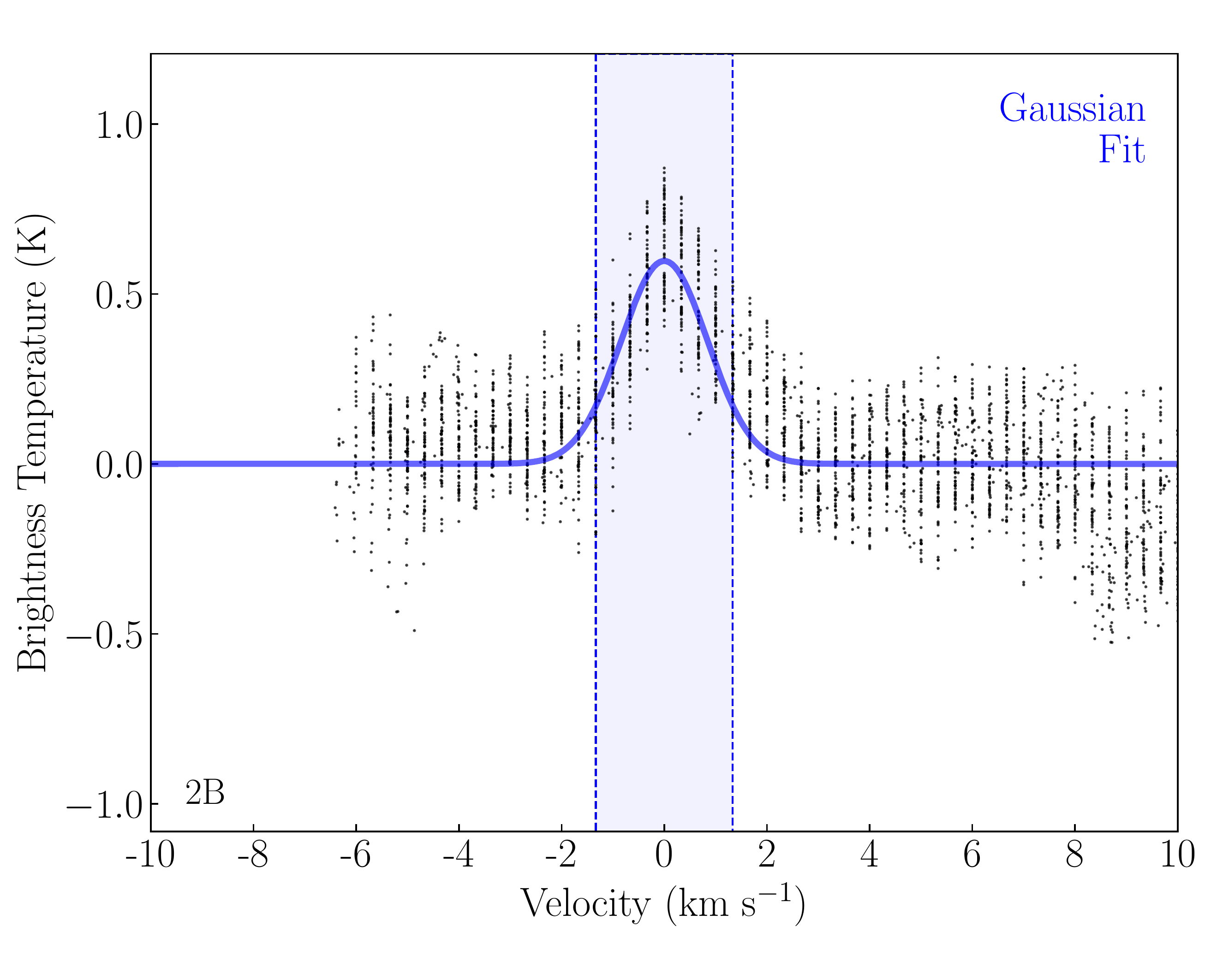}}
\caption{ Velocity profiles generated from the C$^{18}$O $p-v$ diagrams (see Figures \ref{fig:PV_fil_1} and \ref{fig:PV_fil_2}). (\textit{Top Left}) The profile for velocity component 1A centered at approximately -223 km s$^{-1}$. (\textit{Top Right}) The profile for velocity component 1B centered at approximately -208 km s$^{-1}$. (\textit{Bottom Left}) The velocity profile for velocity component 2A centered at approximately -220 km s$^{-1}$.(\textit{Bottom Right}) The profile for velocity component 2B centered at approximately -203 km s$^{-1}$. The blue lines in each figure are the Gaussian fits for each velocity profile made using a fitting distance of 1.3 km s$^{-1}$ from the central velocity. The FWHM velocity linewidths for each velocity component is reported in Table \ref{table:vel_fil_prop}. }
\label{fig:vel_prof}
\end{centering}
\end{figure*}

The velocity profiles with Gaussian fits for each component are given in Figure \ref{fig:vel_prof} and the FWHM velocity linewidths (velocity dispersions) calculated from these fits are reported in Table \ref{table:vel_fil_prop}. We see that both filament 1 and 2 (velocity components 1A and 2B) have velocity linewidths of 2.0 km s$^{-1}$. Filament 1 and 2 both have velocity gradients $<$ 1 km~s$^{-1}$~pc$^{-1}$, with filament 2 having a very small velocity gradient of 0.0017 km~s$^{-1}$~pc$^{-1}$. The larger velocity gradient and linewidth observed in component 2A might be further evidence that it is not associated with gas located along the elongated axis of a coherent filament.

% Velocity Profiles:

\begin{table*}[htb!] 
\centering
\caption{Kinematic properties for the velocity components identified within each filament region.}
%\resizebox{\columnwidth}{!}{
\begin{tabular}{ c c c c } 
\hline\hline
Velocity Component & Central Velocity  & Velocity Gradient  & FWHM Line width \\
 & (km s$^{-1}$) & (km~s$^{-1}$~pc$^{-1}$) & (km s$^{-1}$)  \\
\hline
1A\footnote{Velocity component associated with filament 1.} & -223 & 0.35   & 2.0$\pm{0.28}$ \\ 
1B & -209 & 0.57   & 1.3$\pm{0.13}$ \\  
2A & -219 & 5.2    & 2.4$\pm{0.46}$ \\  
2B\footnote{Velocity component associated with filament 2.} & -203 & 0.0017 & 2.0$\pm{0.19}$ \\ 
\hline
\end{tabular}
%}
\label{table:vel_fil_prop}
\end{table*}

\begin{figure*}[htb!] 
\epsscale{0.55}
\begin{centering}
\subfigure{
\plotone{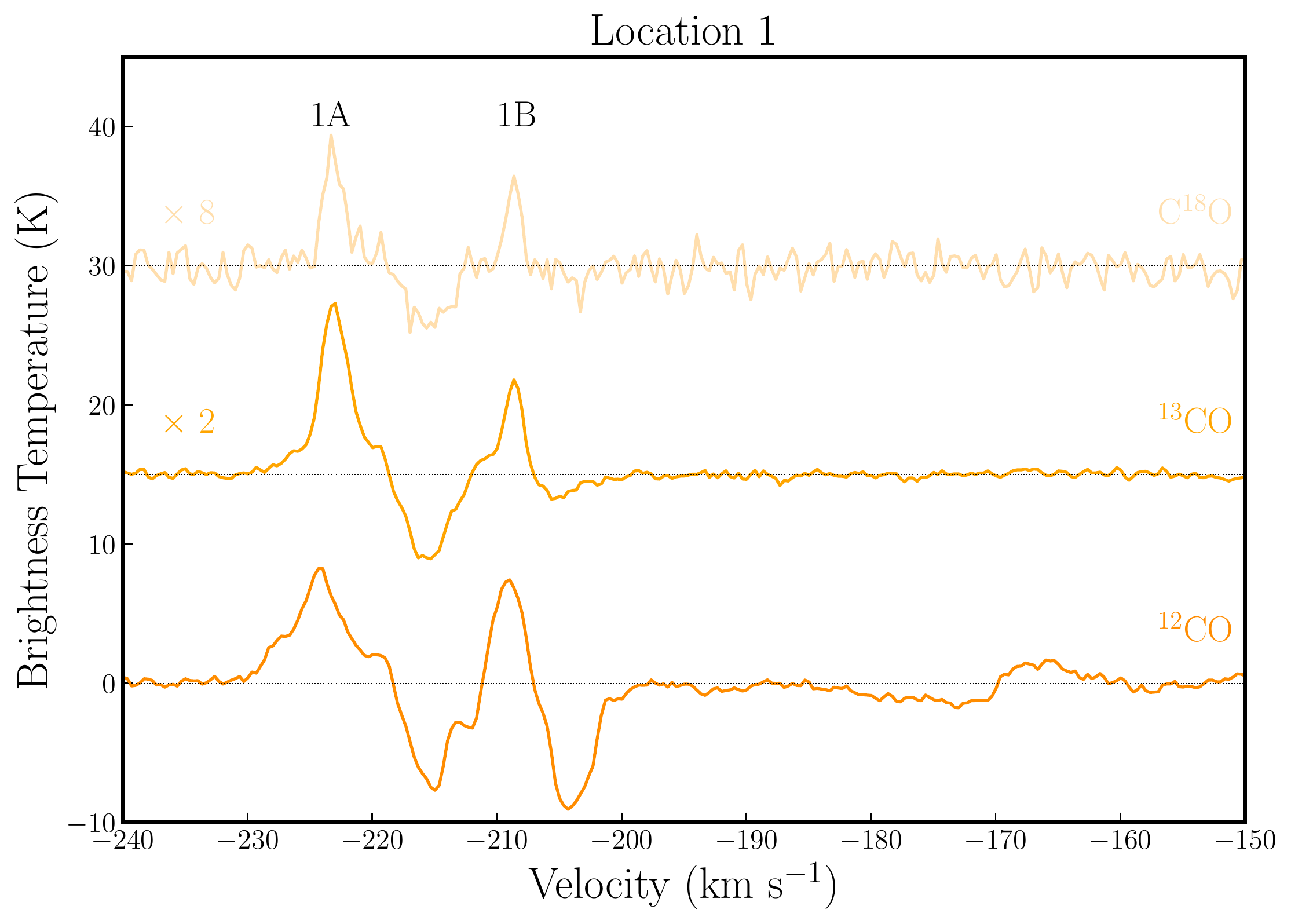}}
\subfigure{
\plotone{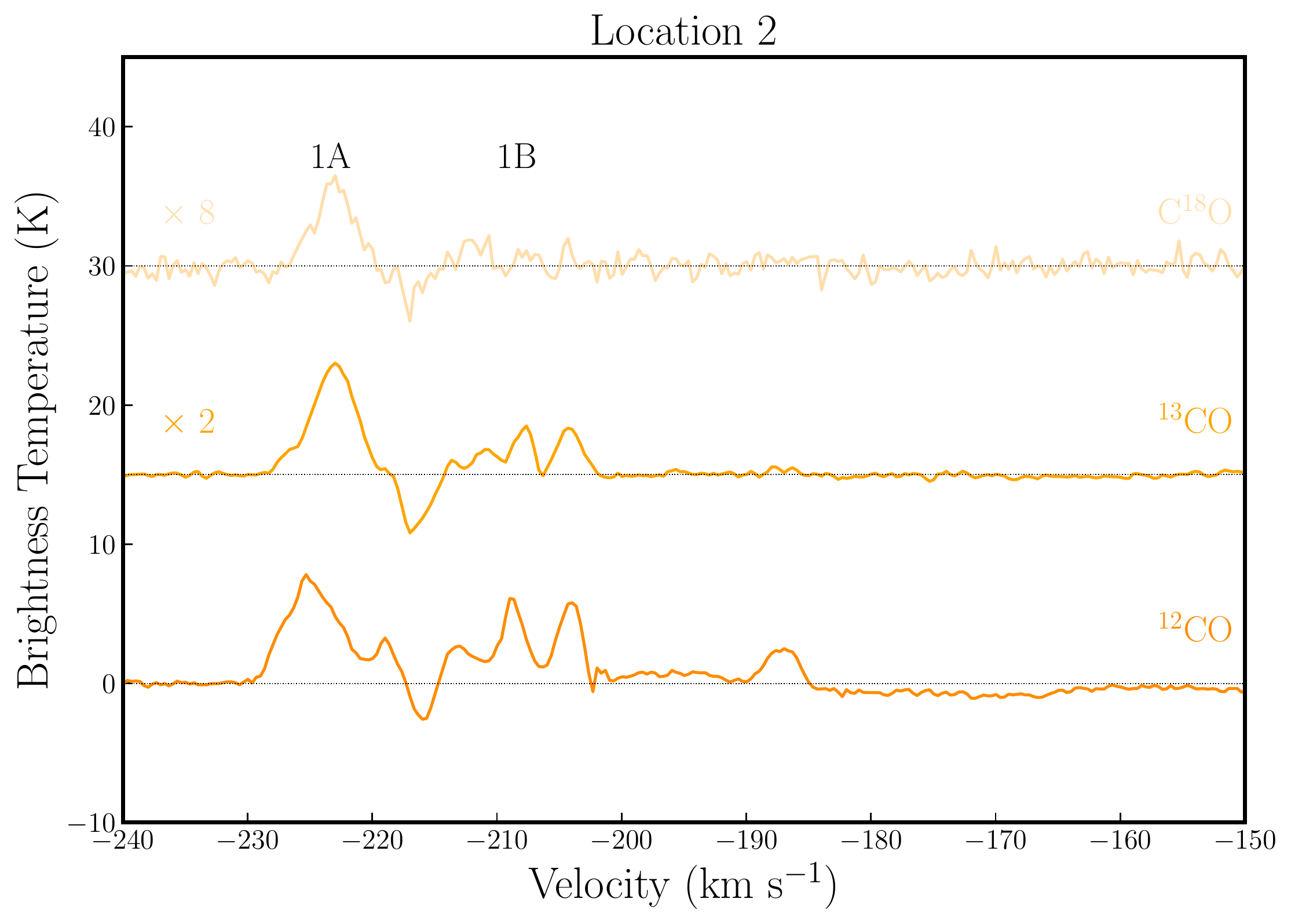}}
\subfigure{
\plotone{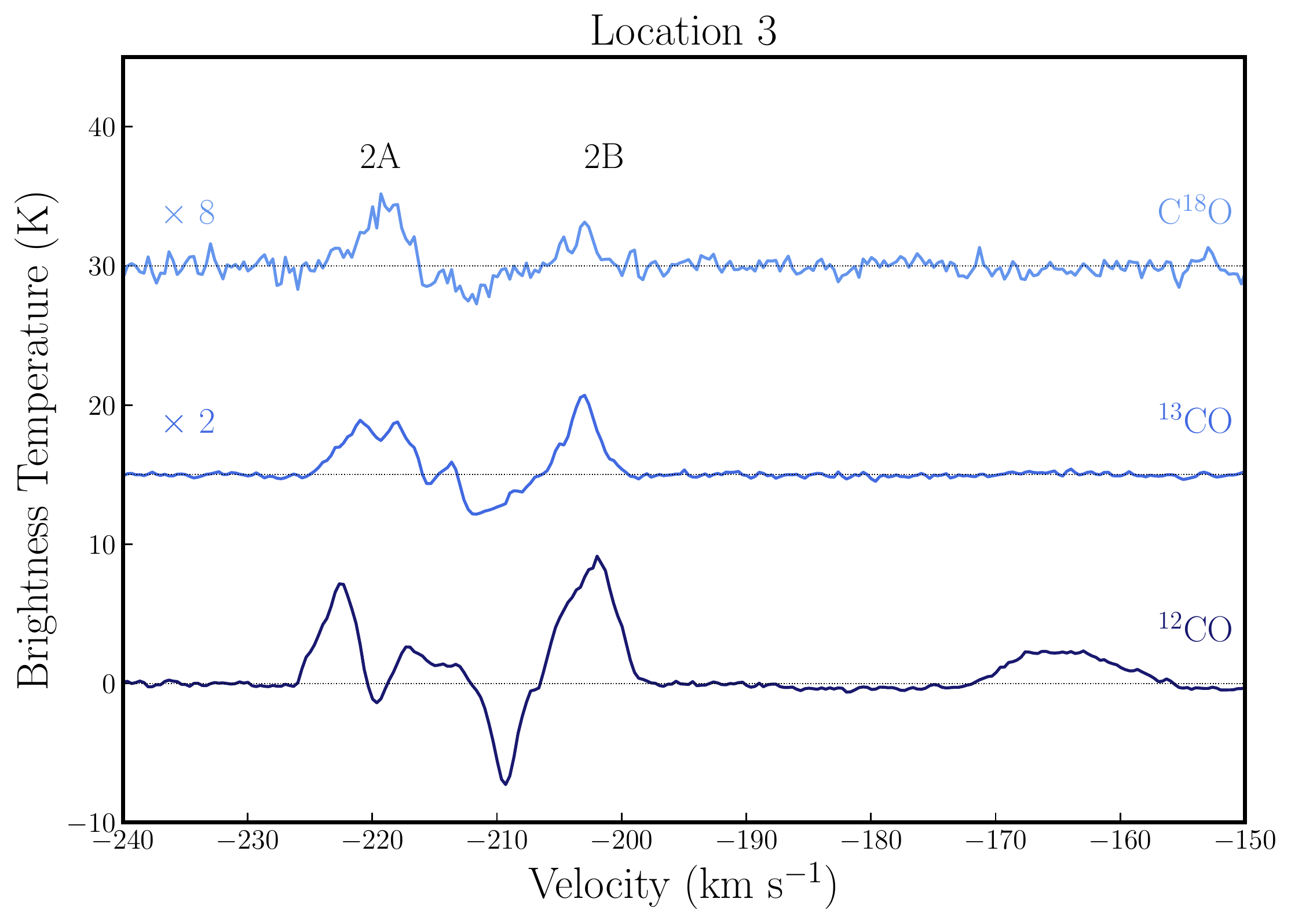}}
\subfigure{
\plotone{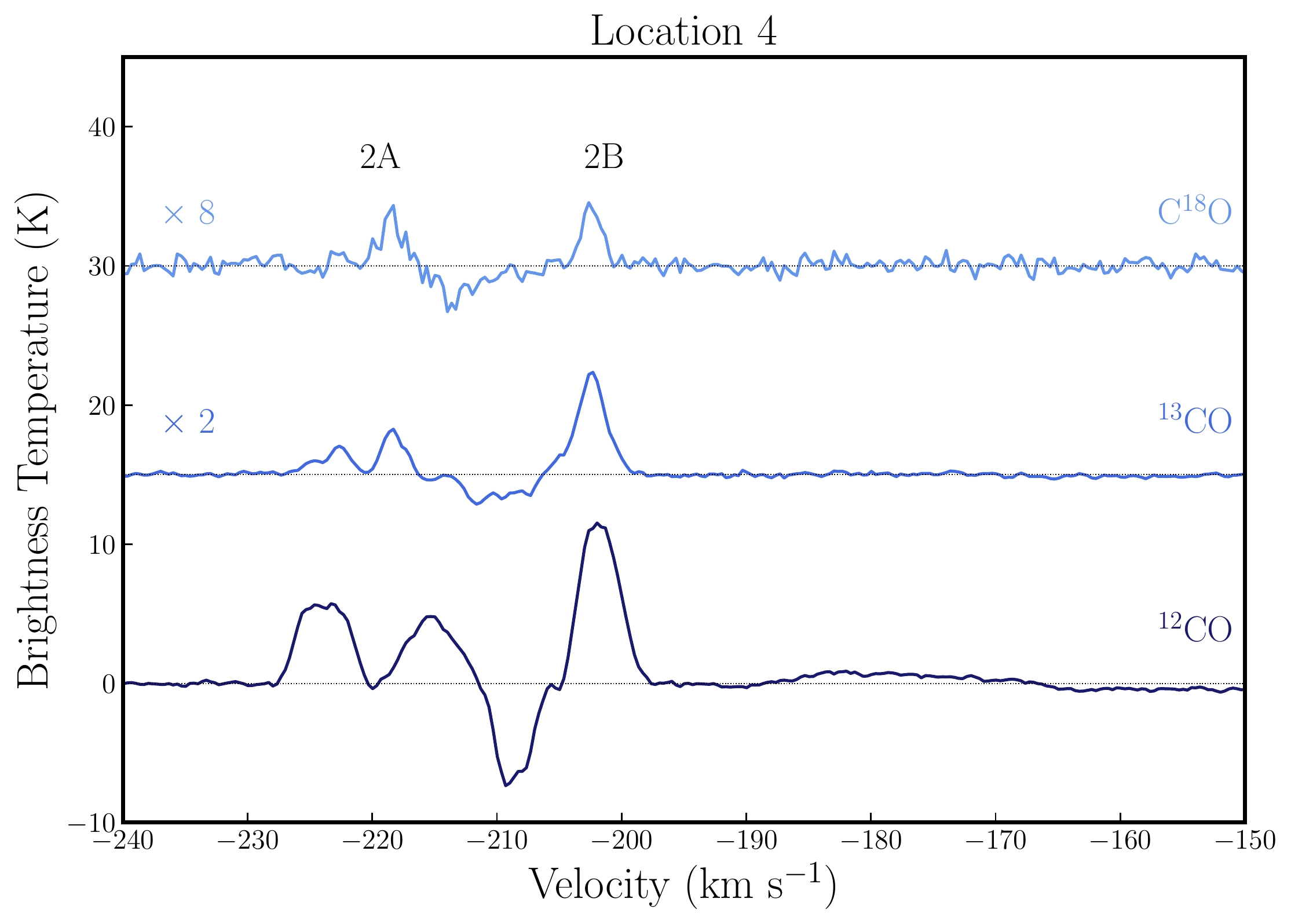}}
\caption{Spectra for $^{12}$CO, $^{13}$CO and C$^{18}$O taken from four different regions and averaged over a circular aperture with a diameter that is three times larger than the semimajor axis of the beam. Regions 1 and 2 were pulled from two different locations along filament 1, whereas regions 3 and 4 were located along filament 2. The spectra for $^{13}$CO and C$^{18}$O are scaled by a factor of 2 and 8 and offset by 15 K and 30 K, respectively, so that they can be seen clearly. Spectral peaks are labeled with their corresponding velocity components identified in the \textit{p-v} diagrams (Figures \ref{fig:PV_fil_1} and \ref{fig:PV_fil_2}).}
\label{fig:avg_spectra}
\end{centering}
\end{figure*}

To further investigate the velocity structure of each filament, we extract spectra at four separate locations along the filaments for all three isotopologues (Figure \ref{fig:avg_spectra}). We chose two sufficiently bright locations on either side of each filament to better represent its full length. In order to improve signal to noise, we averaged spectra in a circular aperture with a diameter three times the semimajor axis of the beam. These apertures are indicated in cyan in Figure \ref{fig:ratio}.

The spectra taken at locations 1 and 2 are given in orange and correspond to the filament 1 region, whereas locations 3 and 4 are given in blue and correspond to the filament 2 region. We denote the spectral peaks using the velocity component labels defined in Section \ref{subsubsection:fil_velocity_analysis}. Overall, we see that the labeled peaks are observed in all three spectral line data. We note that the $^{12}$CO spectrum at location 2  has a peak at approximately -225 km s$^{-1}$, which is displaced from the observed peak at -223 km s$^{-1}$ observed in the $^{13}$CO and C$^{18}$O spectra. Similarly, there is a spectral line component observed at -220 km s$^{-1}$ in the $^{13}$CO and C$^{18}$O spectra at location 4 which is noticeably absent in the $^{12}$CO spectrum. We also note that there are regions of `negative' brightness that appear next to the spectral emission lines which are observed at every location. These are especially exaggerated in the $^{12}$CO emission, but do occur in $^{13}$CO and C$^{18}$O emission as well. These can be interpreted either as regions of self-absorption on a more diffuse CO field, or as an interferometric effect created during the image cleaning process.

% Spectra

\subsection{Effects of opacity}
\label{subsection:opacity}

\begin{figure}[htb!] 
\epsscale{0.98}
\begin{centering}
\subfigure{
\plotone{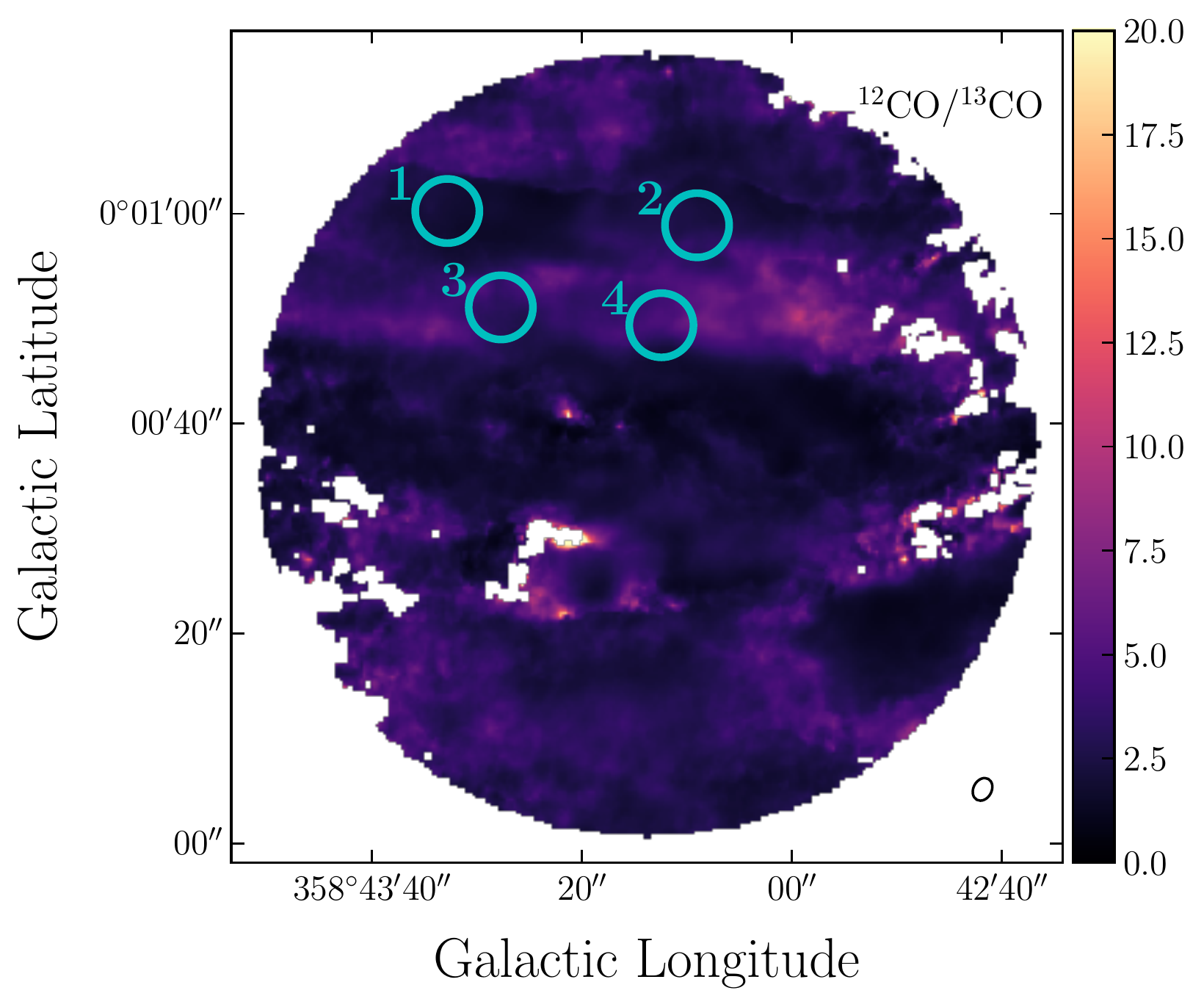}}
\subfigure{
\plotone{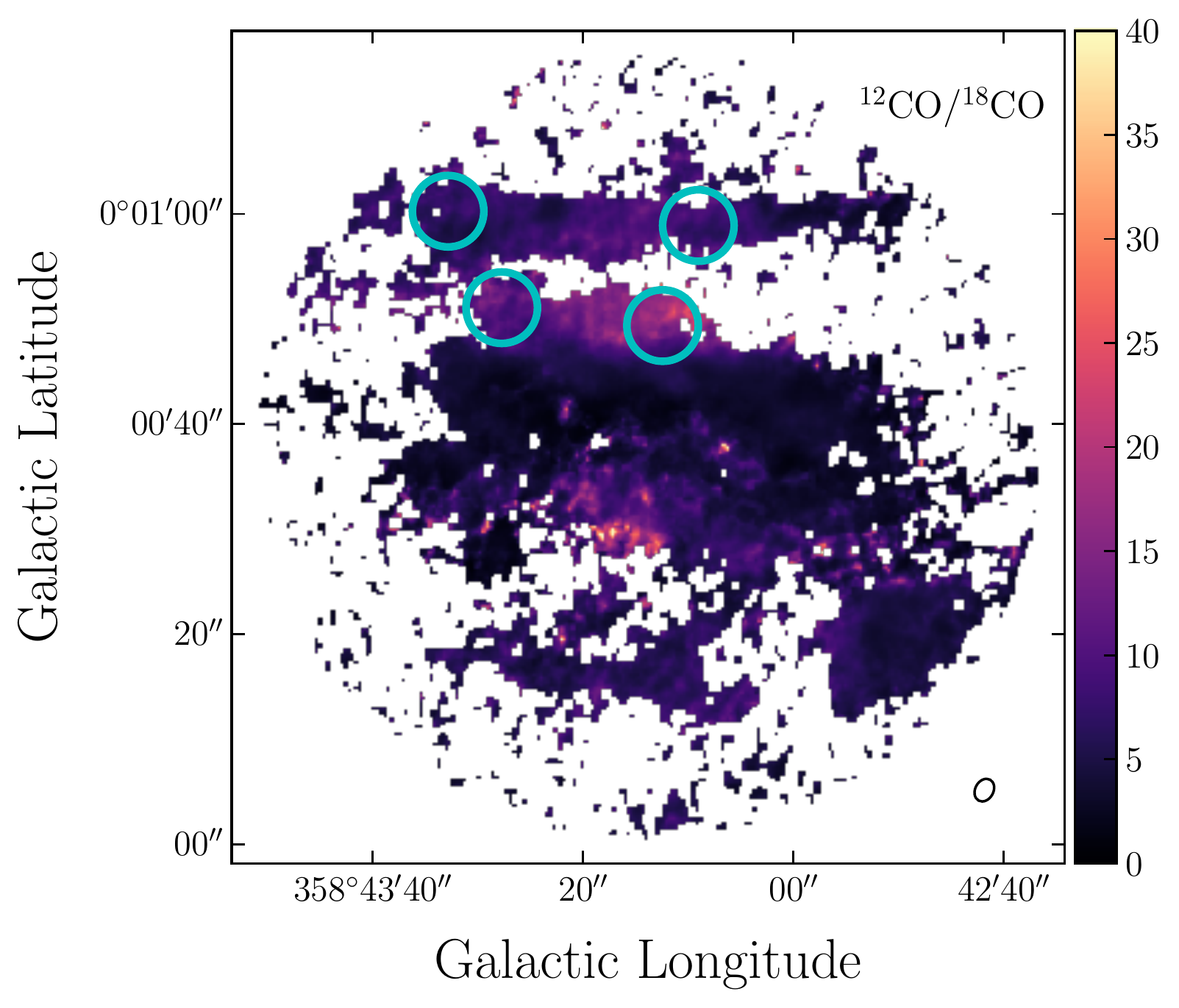}}
\subfigure{
\plotone{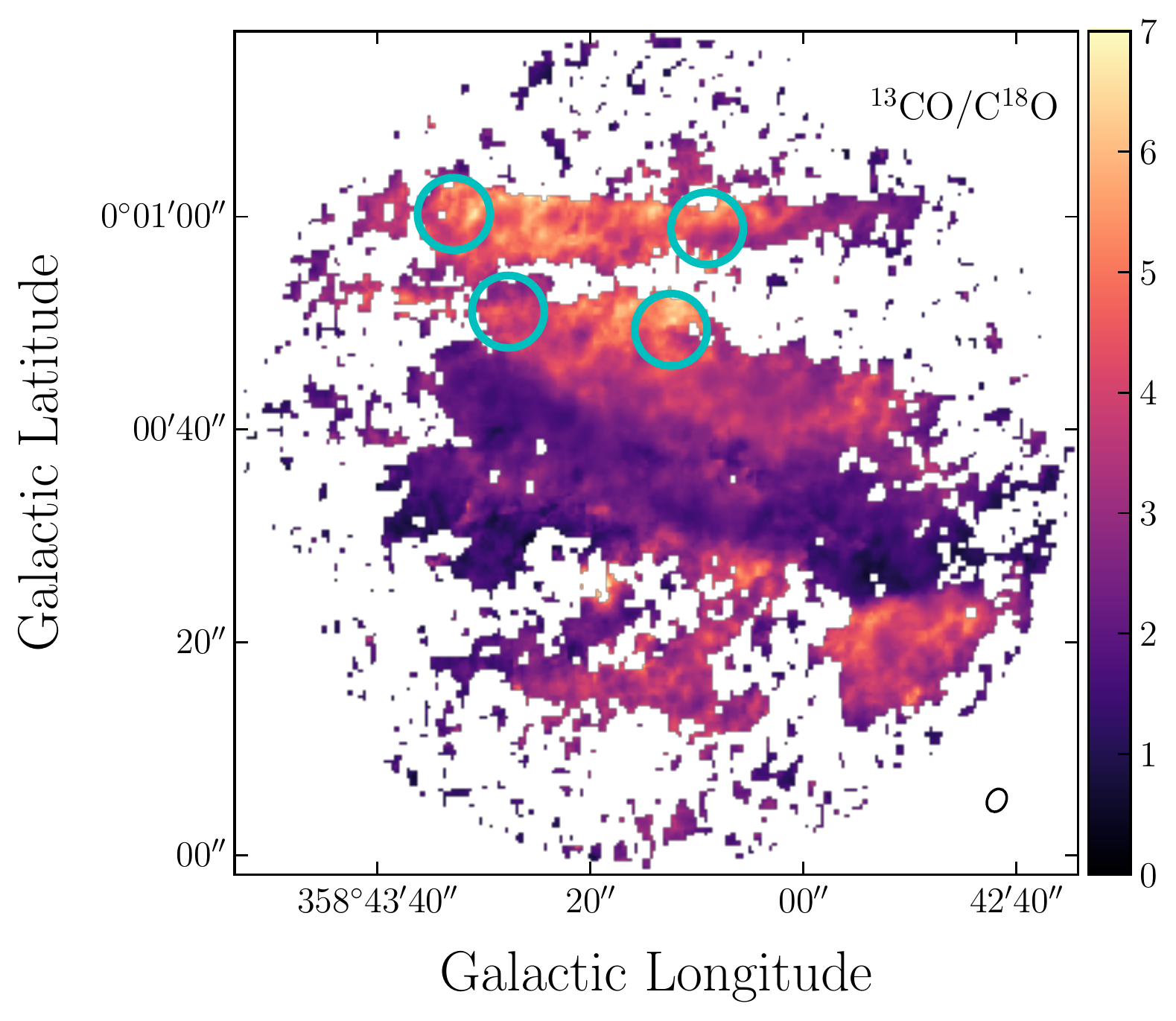}}
\caption{Ratio maps for  $^{12}$C$^{16}$O/$^{13}$C$^{16}$O (\textit{Top}), $^{12}$C$^{16}$O/$^{12}$C$^{18}$O (\textit{Middle}) ,and $^{13}$C$^{16}$O/C$^{18}$O (\textit{Bottom}) . These ratio maps were calculated from the moment 0 maps seen in Figure \ref{fig:mom0} with some additional masking (see Section \ref{subsection:opacity}). The circles outlined in cyan indicate the regions that the spectra were averaged over. In the top figure, the regions are labeled with the number assigned to the spectra in Figure \ref{fig:avg_spectra}.
} 
\label{fig:ratio}
\end{centering}
\end{figure}

It is well known that the spectral lines $^{12}$CO (1-0) and $^{13}$CO (1-0) can be optically thick in observations made towards the Galactic Center where there is a very large amount of molecular gas. In the context of our data set, observations of spectral lines with high opacity may result in the artificial linewidth broadening of the filaments in our FoV. However, since one of our key results is that our linewidths are narrower than expected, this would only serve to make our findings more intriguing. 

To investigate the effects of opacity on our data set, we computed line ratio maps (Figure \ref{fig:ratio}) for $^{12}$C$^{16}$O/$^{13}$C$^{16}$O, $^{12}$C$^{16}$O/$^{12}$C$^{18}$O, and $^{13}$C$^{16}$O/$^{12}$C$^{18}$O, from the integrated intensity maps (Figure \ref{fig:mom0}). We interpret these ratios as indicative of changes in the optical depth across the field. Since the largest angular scale (LAS) is expected to vary by at most 6\% between the $^{12}$CO images and the $^{13}$CO and C$^{18}$O images, and that the structure seen in the $^{13}$CO and C$^{18}$O images does not reach the size scale of the LAS, the effects of interferometric filtering should be negligible, and the interpretation that these ratios reflect changes in the optical depth is reasonable.

To ensure that the same number of pixels were being integrated over for each data cube, we updated the masking so that both moment 0 maps in each ratio calculation were masked identically. To do this, we generated masks for each isotopologue individually as we did for the integrated intensity maps (see Section \ref{subsubsection:moment_analysis}). Then we applied the mask from the other data set being used in the ratio calculation to the first data set. For example, in the $^{12}$CO/$^{13}$CO line ratio calculation, we masked the $^{12}$CO data cube and the $^{13}$CO cube with a combination of both the $^{12}$CO and $^{13}$CO masks that we generated previously. Once all data cubes were masked appropriately, we generated the moment 0 images and directly divided them to produce the desired line ratio maps.

Near the Galactic center, the expected abundance ratio for $^{12}$C/$^{13}$C $\approx$ 20  \citep{wilson_rood_1994,Humire_2020}, whereas the ratio for $^{16}$O/$^{18}$O $\approx$ 250 \citep{wilson_rood_1994}. Given the high column densities and optical depths of our images, we expect that most carbon along the line of sight is in CO and that these isotope ratios are analogous to the corresponding CO isotopologue ratios. 

In our $^{12}$CO/$^{13}$CO line ratio map, we see that most of the FoV exhibits a line ratio less than 10, indicating that $^{12}$CO is optically thick. The filament 1 region exhibits a $^{12}$CO/$^{13}$CO line ratio $< 5$, which suggests that $^{13}$CO is also optically thick in this region. 

Similarly, the line ratios seen in the C$^{16}$O/C$^{18}$O ratio map are far lower than the canonical $\sim$ 250 value. Considering the high opacity of each spectral line, we decided to use the C$^{18}$O data to calculate the filament widths, velocity widths, and velocity gradients, since it is likely the most optically thin line that we observed.

It should be noted that observations were also made for the C$^{17}$O (1-0) spectral line, which is even less abundant than the C$^{18}$O isotopologue \citep[e.g][]{wilson_rood_1994}. Although there is some faint C$^{17}$O emission located at Galactic latitudes south of the observed filaments, there is no clear emission in the filament regions we define. As a result, we cannot provide evidence of optical depth effects on our results using the C$^{17}$O data set.

\begin{figure}[htb!]
    \centering
    \includegraphics[scale=0.45]{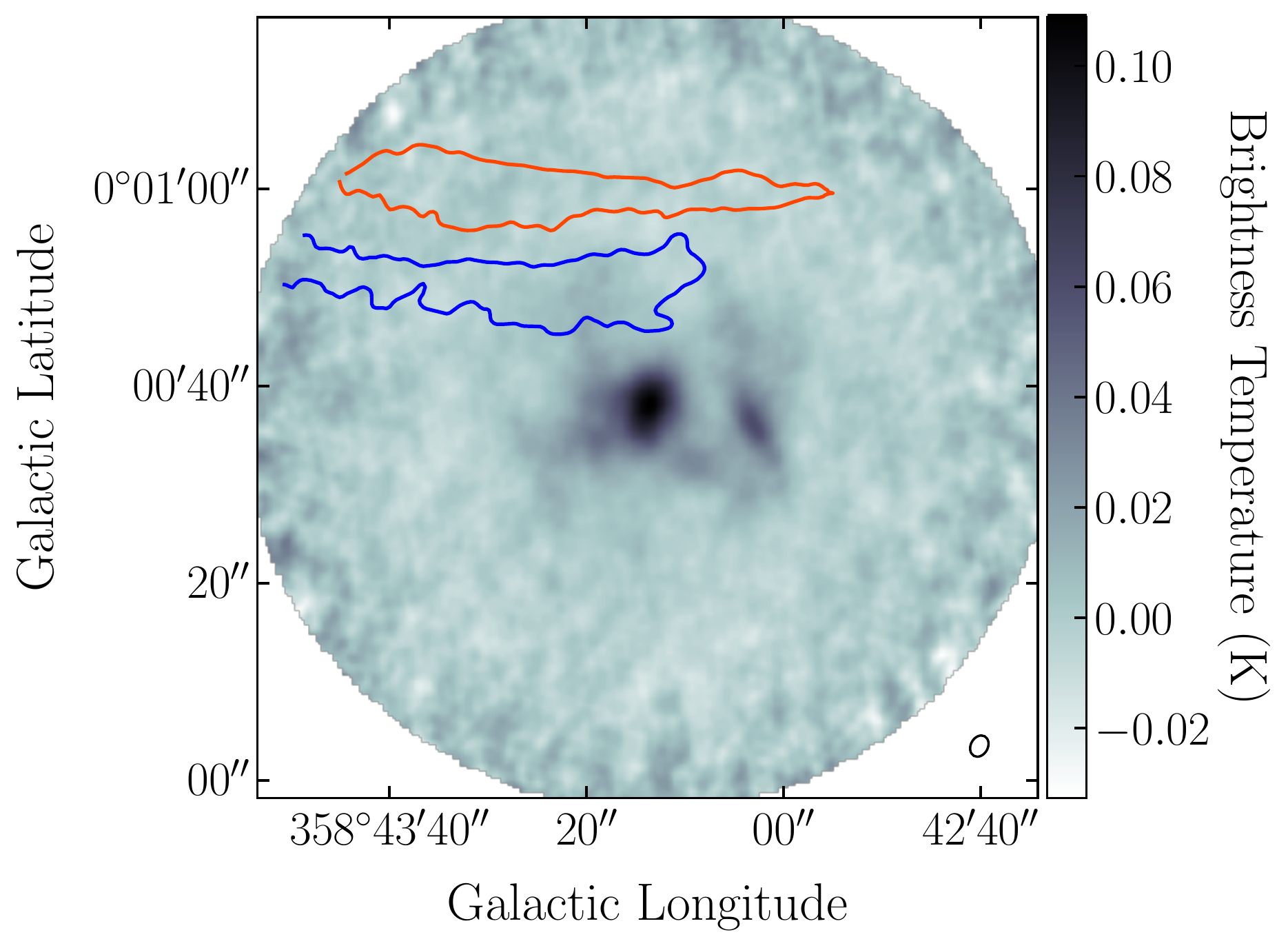}
    \caption{Continuum emission (3 mm) located towards our pointing of the Sgr E region. The top, orange contour indicates the filament 1 region, while the bottom, blue contour indicates the filament 2 region. The beam is indicated in the lower right corner of the figure.}
    \label{fig:continuum}
\end{figure}

\subsection{Dust continuum emission}

In Figure \ref{fig:continuum}, we see that filaments 1 and 2 show no morphological association with the 3mm dust continuum observed at the same pointing. We assume that this continuum emission is primarily from the HII region located at this pointing. The central line-of-sight velocity of the HII region is -206.1 km s$^{-1}$ \citep{Anderson_2020}, compared with central velocities of filament 1 and 2 of -223 km s$^{-1}$ and -203 km s$^{-1}$, respectively. Therefore, though the filaments show no clear morphological association with the continuum emission, they are consistent with being in relatively close proximity in p-p-v space, particularly filament 2.

\section{Discussion}
\label{section:discussion}

\begin{figure}[htb!] 
\centering
\includegraphics[scale=0.53
]{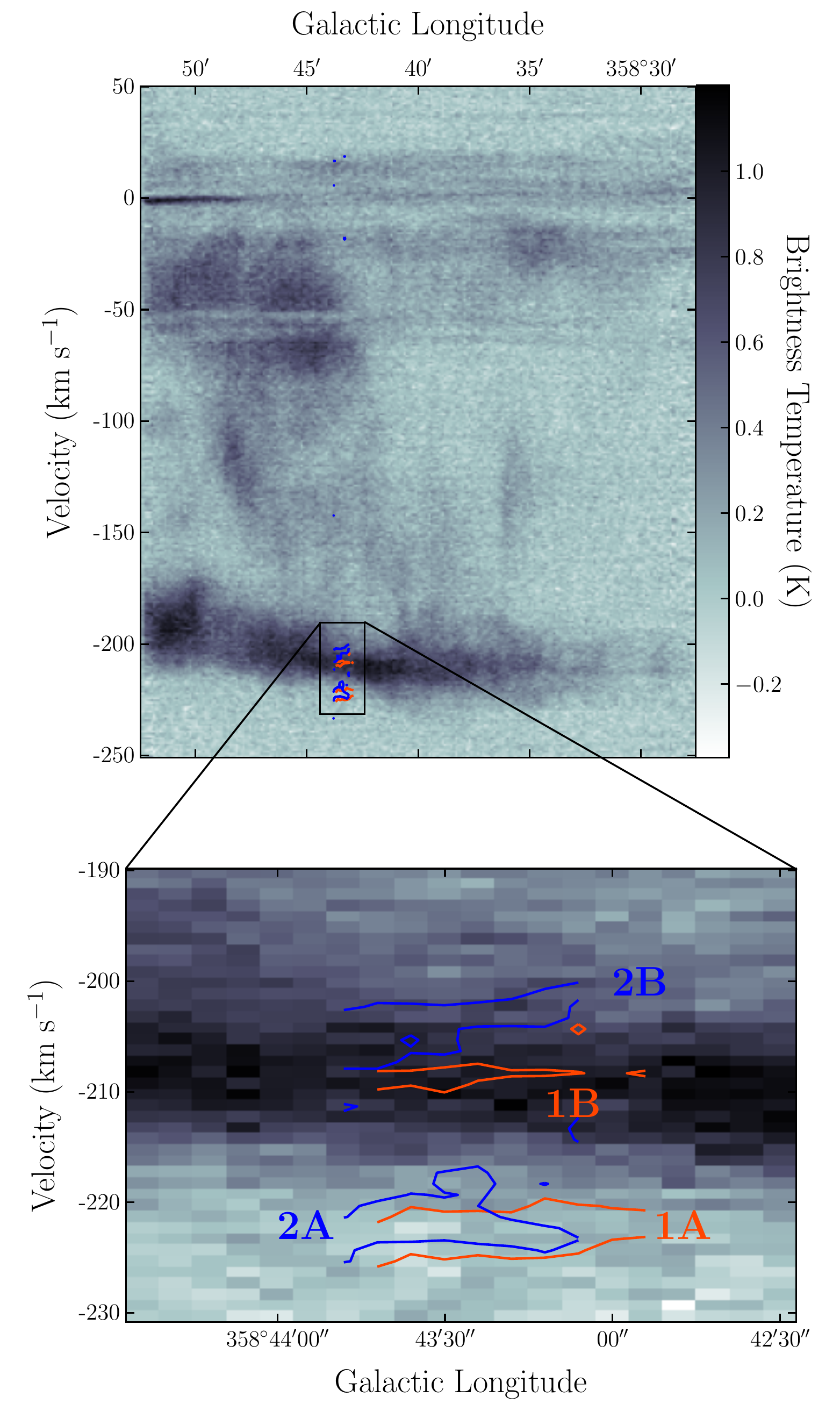}
\caption{
Longitude-velocity diagram of $^{12}$CO (J=3-2) emission \citep[CHIMPS2;][]{Eden_2020} towards the Sgr E region integrated over the range $|$\textit{b}$|$ $<$ 0.175\degree. The data has a spatial resolution of 15\arcsec and a velocity resolution of 1 km s$^{-1}$. The black rectangle represents the boundary of the zoomed-in region shown at the top of the figure. The zoomed-in region spans the velocities -190 km s$^{-1}$ to -230 km s$^{-1}$. The red contours indicate emission from the Sgr E $^{13}$CO data cube in the same longitude-velocity region at a temperature threshold of 0.0075 K. }
\label{fig:overview_lv}
\end{figure}

Our analysis has characterized two prominent filamentary structures towards the Sgr E region and provided information regarding their physical and kinematic properties. We find these filaments to have a linear structure, with small angles towards the Galactic plane. 

In this section, we intend to contextualize the unique features of these structures with respect to the large-scale gas dynamics occurring towards the Sgr E region. To do this, we compare the physical properties of the Sgr E filaments with properties observed in molecular clouds and filaments in the CMZ as well as the Galactic disk. We then discuss the known properties of the HII region associated with our field of view. After this, we provide a discussion of the physical mechanisms that might be creating these filaments. We close this section with open questions that these data present and predictions to be tested in the future.

\subsection{Comparison of Sgr E properties with the CMZ and the Galactic disk}

The Sgr E region is located at the intersection between the far dust lane and the CMZ. The HII regions associated with the Sgr E complex have some of the highest absolute line-of-sight velocities ($\sim$ -200 km s$^{-1}$) of any known star-forming regions in the CMZ or the Galactic disk \citep{Anderson_2014}.  

In Figure \ref{fig:overview_lv}, we provide a longitude-velocity (\textit{l-v}) diagram from the CHIMPS2 $^{12}$CO (3-2) survey where the orange and blue contours represent the $^{13}$CO emission associated with filaments 1 and 2 at a temperature threshold of 0.0075 K. We see that the emission associated with the velocity components 1B and 2B is coincident with the overall large-scale distribution of the CHIMPS2 $^{12}$CO data (Figure \ref{fig:overview_lv}). On the other hand, the emission associated with velocity components 1A and 2A appears to be offset from the main distribution of gas towards more negative velocities by $\sim$ 5 km s$^{-1}$. To check if there was any emission at this velocity detected near the filament regions in the CHIMPS2 data cube, we created an integrated intensity map for velocities ranging from -220 km s$^{-1}$ to -225 km s$^{-1}$. We find that there is $^{12}$CO (3-2) emission located in the region of our filaments for this velocity range, however it is faint compared to nearby emission. We conclude that this emission at $\sim$ -223 km s$^{-1}$ is not apparent in Figure \ref{fig:overview_lv} because it is much fainter than nearby emission, especially when integrated over Galactic latitude.

The length and radial widths of our filaments are well within the expected range for the physical dimensions of filaments in the Galaxy \citep{Hacar_2022}. There is discussion within the literature on the presence of a mean universal filament width of FWHM$\sim$0.1 pc. However, there is evidence that measured filament widths may depend on filament scale and density \citep[][and see references therein]{Hacar_2022}, as well as on the resolution of observations \citep{Panopoulou_2022}. As a result, we conclude that it is likely coincidental that our filaments have FWHM$\sim$0.1 pc, and that this measurement does not suggest the existence of a universal filament width.  

The FWHM linewidth of the filament velocity components reported in Table \ref{table:vel_fil_prop} range between 1.3 - 2.4 km s$^{-1}$. When we convert these linewidths into a total gas velocity dispersion\footnote{See \citet[Section 2]{Hacar_2022} for calculation of $\sigma_{\rm tot}$.}, $\sigma_{\rm tot}$ and divide this by the typical sound speed for CO bright molecular gas ($c_s$ = 0.2 km s$^{-1}$), we find that the velocity dispersion of our filaments, in units of sound speed, are between $\frac{\sigma_{\rm tot}}{c_s}$ = 2.3 - 3.9. These values are well within the range that is observed for other Galactic filaments with lengths of $\sim$ 2 pc, approximately $\frac{\sigma_{\rm tot}}{c_s}$ = 1 - 10  \citep{Hacar_2022}.

The velocity dispersion of molecular gas structures in the CMZ tend to be much larger than those observed in the disk \citep{Shetty_2012,Henshaw_2016}.  At a similar sub-parsec size scale, our FWHM linewidths are narrower than those measured in dense star forming cores in the CMZ, with velocity dispersions from 1--4 km s$^{-1}$, corresponding to FWHM linewidths of $\sim$ 2--9 km s$^{-1}$ \citep{Callanan_2021}. 

Interestingly, the velocity linewidths observed in our filaments seem to be similar to filaments in the Galactic disk, and somewhat narrow when compared to gas in the CMZ. The larger linewidths for gas in the CMZ are thought to be driven mostly by turbulence \citep{Shetty_2012, Kauffmann_2017} which may be caused in part by collisions from gas streaming into the CMZ from the dust lanes \citep{Henshaw_2022}. Based on our observations, the gas in the Sgr E region appears to be relatively quiescent when compared to gas within the CMZ as it travels along the far dust lane.

\subsection{Discussion of the origin of Sgr E's filamentary structure.}

Using high resolution CO spectral line data, we have shown that the Sgr E region contains multiple filaments that exhibit a linear structure with an orientation that is nearly parallel to the Galactic plane. These filaments also possess an interesting kinematic structure, with velocity components along the line of sight that have comparatively narrow velocity linewidth values.

We propose that the properties of the Sgr E filaments may be due to their unique placement between the CMZ and the far dust lane. As previously mentioned, the molecular gas in the Sgr E region is located where the far dust lane meets the CMZ, and moves with a high line-of-sight velocity of $\sim$ -200 km s$^{-1}$. There is evidence which suggests that the location of the Sgr E region may have caused it to develop some unique infrared properties. For example, \cite{Anderson_2020} measured unusually high 22 $\mu$m to 12 $\mu$m flux density ratios towards many of the Sgr E HII regions; a possible indication of a lack of photodissociation regions (PDRs) in the complex. They proposed that the interaction between the high velocity Sgr E HII regions and the local gas could possibly strip the regions of their PDRs.

\begin{figure}[htb!] 
\centering
\includegraphics[scale=0.4]{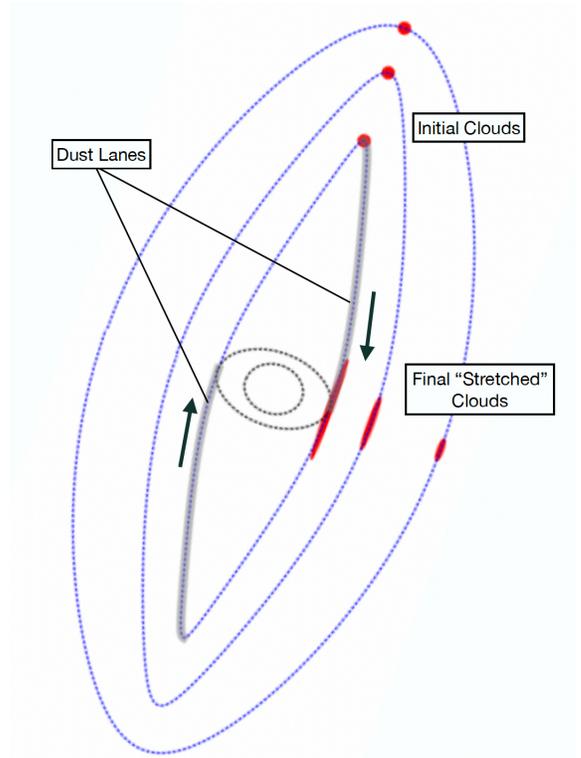}
\caption{A cartoon image indicating the stretching of molecular gas as it travels along x$_1$ orbits, from apocenter to pericenter. The red circles represent the initial parcels of gas located at apocenter and the red ellipses represent their elongated counterparts once they reach pericenter. The blue dotted lines indicate various x$_1$ orbits and the black dotted lines indicate x$_2$ orbits. The shaded gray regions indicate possible dust lane placements. The black arrows show the direction of gas flow. The x$_1$ and x$_2$ orbits and the factor by which gas is stretched was calculated using the same gravitational potential used in \cite{Sormani_2018} (their orbits are shown in the top-left panel of their Figure 8).
}
\label{fig:cartoon}
\end{figure}

Similarly, we suggest that the Sgr E filaments may be products of their unique location at the intersection of the CMZ and the far dust lane. High resolution numerical simulations of disk galaxies demonstrate that filamentary structures can be created in the disk both by Galactic dynamical effects such as differential rotation  and also by stellar feedback \citep{Smith_2020}. However, filaments created by feedback are as likely to be found perpendicular to the Galactic plane as parallel to it. The high degree of alignment between the Sgr E filaments and the galactic plane therefore leads us to favor a dynamical origin for them. The orientation of the filaments, as well as their linear structure could indicate that the molecular gas is strongly influenced by the Galactic bar potential, which drives the flow of gas along the dust lanes. This strong gravitational influence may be `stretching' the molecular gas of the Sgr E region along the Galactic plane and towards the Galactic center. The dust lanes are composed of gas on nearly radial orbits, meaning that the gas is subject to large accelerations parallel to the direction of motion. Since the gas in Sgr E is near the pericenter of very elongated x$_1$-type orbits, it should be at a point of ``maximum-stretching". The cartoon in Figure \ref{fig:cartoon} shows how gas can be stretched along these nearly radial orbits. For example, if a spherical molecular cloud is placed at the tip of the far dust lane ($\sim$ 3 kpc from the Galactic center), by the time it reaches the position of Sgr E, it will resemble the shape of an ellipse. This effect would be most dramatic along the Galactic bar dust lanes (highlighted in gray in Figure \ref{fig:cartoon}) as these features are closely aligned with the innermost x$_1$ orbits. For the innermost x$_1$ orbit shown in Figure \ref{fig:cartoon}, which is calculated using the same gravitational potential used for the gas flow simulations in \cite{Sormani_2018}, the stretching factor (i.e. the axis ratio of the cloud at pericenter assuming it starts spherical at apocenter) is about $\sim$10. We stress that this proposed origin for the Sgr E filaments is one plausible explanation based on the limited data set and prior information, and that a more thorough investigation of the region is needed to make a stronger statement about the formation of these structures.

\subsection{Open questions and future directions}

Based on the analysis presented in this paper, we speculate that these filaments formed due to accelerations imposed by the Galaxy bar potential. To fully test this prediction, it will be important to determine if these filamentary structures are observed elsewhere in the Sgr E complex. This could be done by completing multiple observations surveying a larger portion of the Sgr E region at a comparable spatial resolution. In addition to this, it would be useful to expand the target region to see the full length of the filaments. Another strong test of our dynamical origin prediction would be to make similar observations towards the ending portion of the near dust lane. Observing a pervasive population of molecular filaments with orientations and physical properties similar to the filaments reported on in this paper would support our hypothesis that the molecular gas in the Sgr E region is being stretched into filaments along its orbital path due to the gravitational influence from the Galactic bar. It would also be informative to compare these observations with hydrodynamical simulations of comparable resolution to see if such filamentary structures are formed in a region analogous to the location of Sgr E.

\section{Conclusion}
\label{section:conclusion}
Sgr E is a unique and dynamic HII region complex situated at the intersection between the CMZ and the far dust lane. With our analysis of new ALMA data, we have shown that:

\begin{enumerate}
    \item We observe two prominent, distinct filaments in p-p-v space within the small part of the Sgr E region targeted in our ALMA observations. These filaments are seen clearly in $^{12}$CO, $^{13}$CO and C$^{18}$O emission and we refer to them as filaments 1 and 2.
    \item These two filaments have measured lengths of at least $\sim$ 2 pc and FWHM widths of $\sim$ 0.1 pc.
    \item Both filaments are aligned nearly parallel to the Galactic plane, with position angles $\sim$ 2\degree.
    \item Filaments 1 and 2 have central velocities of -223 km s$^{-1}$ and -203 km s$^{-1}$, respectively.
    \item There are two other line-of-sight velocity components centered at -209 km s$^{-1}$ and -219 km s$^{-1}$ that are observed along the lengths of filament 1 and 2, respectively. We determine that these components are not directly associated with filaments 1 and 2, but are likely associated with either local non-filamentary emission in the field of view or a separate coherent structure that is co-spatial with the identified filament regions.
    \item Filaments 1 and 2 have velocity gradients $<1$ km~s$^{-1}$~pc$^{-1}$, within the expected range for filaments found in typical Galactic environments.
    \item Both filaments have FWHM velocity linewidths of $\sim$2.0 km s$^{-1}$. These linewidths are narrow when compared to those measured in the CMZ, and are similar to those measured in the Galactic disk on comparable scales.
\end{enumerate}

We propose that the physical and kinematic properties of these filaments can be explained by considering the location and dynamics of the Sgr E region with respect to the Galaxy. The elongation and orientation of our filaments might be caused by the ``stretching" of molecular gas in Sgr E due to the gravitational influence of the Galactic bar. Further investigation is needed to support or refute this line of reasoning. It would be useful to observe CO emission towards other pointings in the Sgr E complex and in the analogous portion of the near dust lane at a similar resolution and with a larger field of view. Identifying other filaments with similar properties to those reported here would provide important insight on the larger scale gas motions that occur between the Galactic bar dust lanes and the CMZ, and how that affects the small scale gas motions in this region.\\
\\
\\
\\
\noindent 
JW gratefully acknowledges funding from the National Science Foundation under Award No. 2108938.

CB gratefully  acknowledges funding  from  National  Science  Foundation  under  Award  Nos. 1816715, 2108938 and CAREER 2145689, as well as from the National Aeronatics and Space Administration through the Astrophysics Data Analaysis Program under Award No. 21-ADAP21-0179 and through the SOFIA archival research program under Award No. 09$\_$0540.  

Financial support for this work was provided by NASA through award $\#$09$\_$0540 issued by USRA. 

Support for this work was provided by the NSF through award SOSPA7-007 from the NRAO.

MCS acknowledges financial support from the European Research Council via the ERC Synergy Grant ``ECOGAL‚ Äì Understanding our Galactic ecosystem: from the disk of the Milky Way to the formation sites of stars and planets'' (grant 855130).

ATB would like to acknowledge funding from the European Research Council (ERC) under the European Union’s Horizon 2020 research and innovation programme (grant agreement No.726384/Empire).

SCOG acknowledges support from the Deutsche Forschungsgemeinschaft (DFG) via SFB 881 ``The Milky Way System'' (Project-ID 138713538; sub-projects B1, B2 and B8) and from the Heidelberg cluster of excellence EXC 2181-390900948 ``STRUCTURES: A unifying approach to emergent phenomena in the physical world, mathematics, and complex data'', funded by the German Excellence Strategy. He also acknowledges funding from the ERC via the ERC Synergy Grant ECOGAL (grant 855130).

AG acknowledges support from the NSF under awards AST 2008101 and and CAREER 2142300.

The National Radio Astronomy Observatory is a facility of the National Science Foundation operated under cooperative agreement by Associated Universities, Inc.

This paper makes use of the following ALMA data: ADS/JAO.ALMA\#2019.1.01240.S. ALMA is a partnership of ESO (representing its member states), NSF (USA) and NINS (Japan), together with NRC (Canada), MOST and ASIAA (Taiwan), and KASI (Republic of Korea), in cooperation with the Republic of Chile. The Joint ALMA Observatory is operated by ESO, AUI/NRAO and NAOJ.
%\end{acknowledgments}

\vspace{5mm}
\facility{ALMA}

\software{Astropy \citep{Astropy_2013, Astropy_2018}, Spectralcube \citep{adam_ginsburg_2019_3558614}, Radfil \citep{Zucker_2018}, Filfinder \citep{2015MNRAS.452.3435K}, pvextractor \citep{ginsburg_pv_2016}, CASA: version 5.6.1-8 \citep{McMullin_2007}}

\appendix

\section{Velocity channel maps for the spectral lines $^{12}$CO and C$^{18}$O}
\label{section:appendix_channels}

\begin{figure*}[htb!] 
\epsscale{1.15}
\begin{centering}
\subfigure{
\plotone{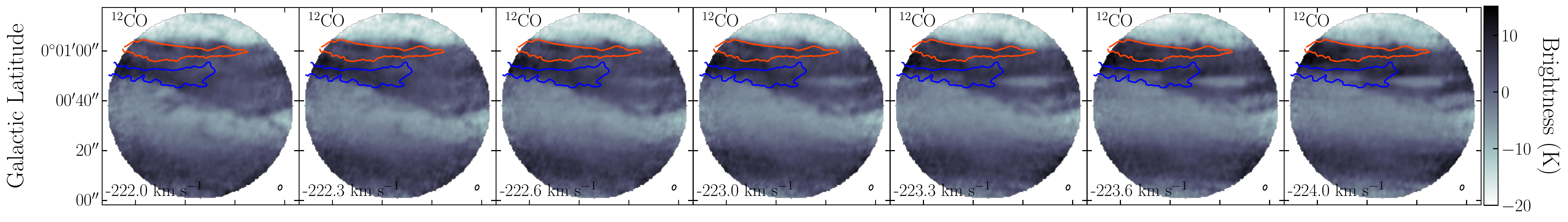}}
\subfigure{
\plotone{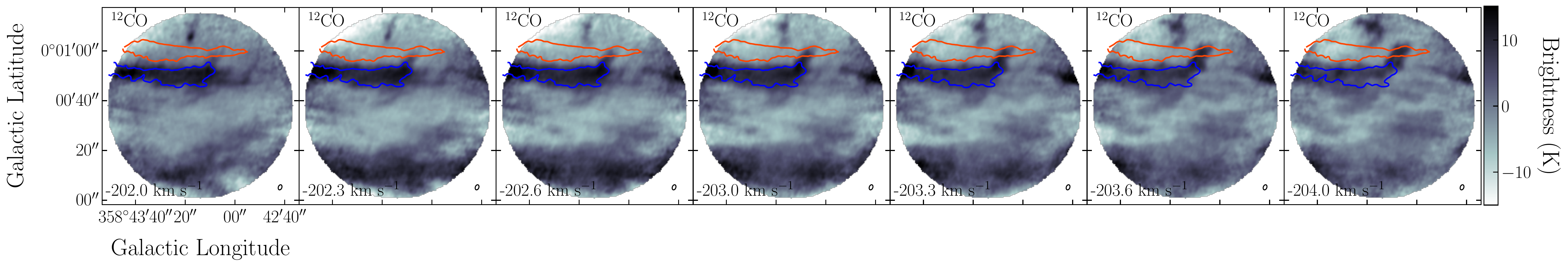}}
\caption{The full resolution velocity channel maps for the spectral line $^{12}$CO with channel widths of $\sim$ 0.3 km s$^{-1}$. The channel maps on the top row span the velocity range of -222.0 km s$^{-1}$ to -224.0 km s$^{-1}$. The channel maps on the bottom row span the velocity range of -202.0 km s$^{-1}$ to -204.0 km s$^{-1}$. In each figure, the top, orange contour indicates the region we define for filament 1, while the bottom, blue contour indicates the filament 2 region. The beam is indicated in the lower right corner of each channel map. }
\label{fig:chan12_fil}
\end{centering}
\end{figure*}

\begin{figure*}[htb!] 
\epsscale{1.15}
\begin{centering}
\subfigure{
\plotone{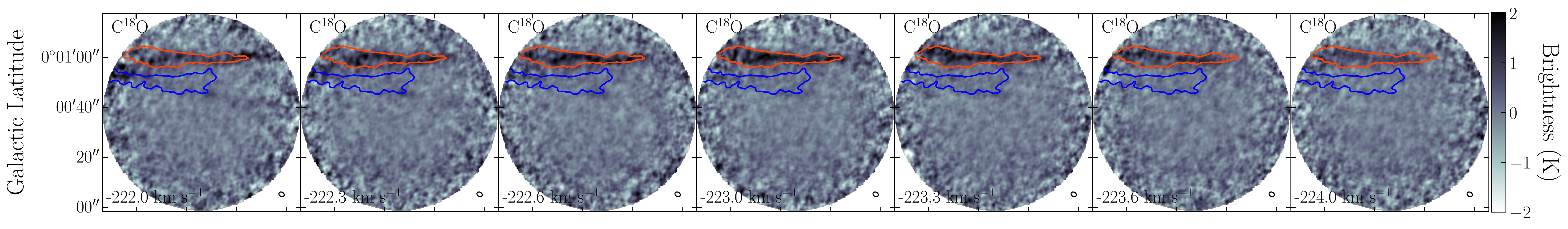}}
\subfigure{
\plotone{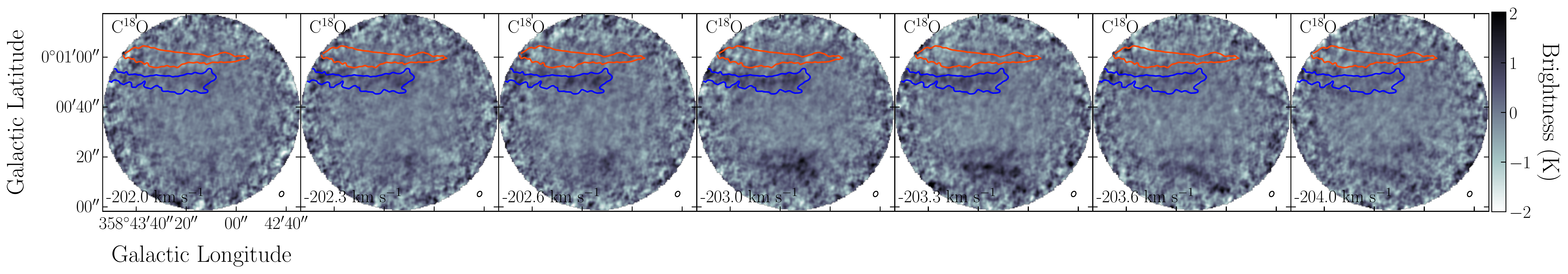}}
\caption{The full resolution velocity channel maps for the spectral line C$^{18}$O with channel widths of $\sim$ 0.3 km s$^{-1}$. The channel maps on the top row span the velocity range of -222.0 km s$^{-1}$ to -224.0 km s$^{-1}$. The channel maps on the bottom row span the velocity range of -202.0 km s$^{-1}$ to -204.0 km s$^{-1}$. In each figure, the top, orange contour indicates the region we define for filament 1, while the bottom, blue contour indicates the filament 2 region. The beam is indicated in the lower right corner of each channel map. }
\label{fig:chan18_fil}
\end{centering}
\end{figure*}

For completeness, we include the full-resolution velocity channel maps for the spectral lines $^{12}$CO and C$^{18}$O (Figures \ref{fig:chan12_fil} and \ref{fig:chan18_fil}). These channel maps have the same channel widths ($\sim$0.3 km~s$^{-1}$) and cover the same velocity ranges as the $^{13}$CO channel maps presented in Figure \ref{fig:chan13_fil}.

% 12CO Velocity channel maps centered on filaments 1 and 2:

% 13CO Velocity channel maps centered on filaments 1 and 2:

\section{Estimation of uncertainty}
\label{section:uncertainty}

The primary source of error for the filament width and velocity linewidth calculations arise from the choice of a fitting distance and background subtraction radius when generating a Gaussian fit. The \verb|RadFil| Python package has a built-in method for estimating this systematic uncertainty which is described in detail in \cite{Zucker_2018}. In short, this method requires that the user inputs a range of fitting distances and background subtraction radii. These inputs then allow \verb|Radfil| to calculate the best fit to the data using all possible combinations. The overall systematic error is then calculated by taking the standard deviation of the resulting ensemble of best-fit parameters. 

For Gaussian fits of the filament profiles, we used fitting distances of 0.025 pc, 0.050 pc, 0.075 pc, and 0.10 pc, and background subtraction radii with an inner bound of 0.25 pc and outer bounds of 0.6 pc, 0.8 pc, 1.0 pc, and 1.2 pc. For Gaussian fits of the velocity component profiles, we used fitting distances of 1.0 km s$^{-1}$, 1.3 km s$^{-1}$, 1.7 km s$^{-1}$, and 2.0 km s$^{-1}$, and background subtraction radii with an inner bound of 2.7 km s$^{-1}$, and outer bounds of 6.7 km s$^{-1}$, 10.0 km s$^{-1}$, 13.3 km s$^{-1}$, and 20.0 km s$^{-1}$. The systematic uncertainties for the filament widths and the velocity component linewidths are reported in Tables \ref{table:fil_prop} and \ref{table:vel_fil_prop}, respectively.


\bibliography{references}{}

\begin{thebibliography}{}
\expandafter\ifx\csname natexlab\endcsname\relax\def\natexlab#1{#1}\fi
\providecommand{\url}[1]{\href{#1}{#1}}
\providecommand{\dodoi}[1]{doi:~\href{http://doi.org/#1}{\nolinkurl{#1}}}
\providecommand{\doeprint}[1]{\href{http://ascl.net/#1}{\nolinkurl{http://ascl.net/#1}}}
\providecommand{\doarXiv}[1]{\href{https://arxiv.org/abs/#1}{\nolinkurl{https://arxiv.org/abs/#1}}}

\bibitem[{{Anderson} {et~al.}(2014){Anderson}, {Bania}, {Balser}, {Cunningham},
  {Wenger}, {Johnstone}, \& {Armentrout}}]{Anderson_2014}
{Anderson}, L.~D., {Bania}, T.~M., {Balser}, D.~S., {et~al.} 2014, \apjs, 212,
  1, \dodoi{10.1088/0067-0049/212/1/1}

\bibitem[{Anderson {et~al.}(2020)Anderson, Sormani, Ginsburg, Glover, Heywood,
  Rammala, Schuller, Csengeri, Urquhart, \& Bronfman}]{Anderson_2020}
Anderson, L.~D., Sormani, M.~C., Ginsburg, A., {et~al.} 2020, \apj, 901, 51,
  \dodoi{10.3847/1538-4357/abadf6}

\bibitem[{{Astropy Collaboration} {et~al.}(2013){Astropy Collaboration},
  {Robitaille}, {Tollerud}, {Greenfield}, {Droettboom}, {Bray}, {Aldcroft},
  {Davis}, {Ginsburg}, {Price-Whelan}, {Kerzendorf}, {Conley}, {Crighton},
  {Barbary}, {Muna}, {Ferguson}, {Grollier}, {Parikh}, {Nair}, {Unther},
  {Deil}, {Woillez}, {Conseil}, {Kramer}, {Turner}, {Singer}, {Fox}, {Weaver},
  {Zabalza}, {Edwards}, {Azalee Bostroem}, {Burke}, {Casey}, {Crawford},
  {Dencheva}, {Ely}, {Jenness}, {Labrie}, {Lim}, {Pierfederici}, {Pontzen},
  {Ptak}, {Refsdal}, {Servillat}, \& {Streicher}}]{Astropy_2013}
{Astropy Collaboration}, {Robitaille}, T.~P., {Tollerud}, E.~J., {et~al.} 2013,
  \aap, 558, A33, \dodoi{10.1051/0004-6361/201322068}

\bibitem[{{Astropy Collaboration} {et~al.}(2018){Astropy Collaboration},
  {Price-Whelan}, {Sip{\H{o}}cz}, {G{\"u}nther}, {Lim}, {Crawford}, {Conseil},
  {Shupe}, {Craig}, {Dencheva}, {Ginsburg}, {VanderPlas}, {Bradley},
  {P{\'e}rez-Su{\'a}rez}, {de Val-Borro}, {Aldcroft}, {Cruz}, {Robitaille},
  {Tollerud}, {Ardelean}, {Babej}, {Bach}, {Bachetti}, {Bakanov}, {Bamford},
  {Barentsen}, {Barmby}, {Baumbach}, {Berry}, {Biscani}, {Boquien}, {Bostroem},
  {Bouma}, {Brammer}, {Bray}, {Breytenbach}, {Buddelmeijer}, {Burke},
  {Calderone}, {Cano Rodr{\'\i}guez}, {Cara}, {Cardoso}, {Cheedella}, {Copin},
  {Corrales}, {Crichton}, {D'Avella}, {Deil}, {Depagne}, {Dietrich}, {Donath},
  {Droettboom}, {Earl}, {Erben}, {Fabbro}, {Ferreira}, {Finethy}, {Fox},
  {Garrison}, {Gibbons}, {Goldstein}, {Gommers}, {Greco}, {Greenfield},
  {Groener}, {Grollier}, {Hagen}, {Hirst}, {Homeier}, {Horton}, {Hosseinzadeh},
  {Hu}, {Hunkeler}, {Ivezi{\'c}}, {Jain}, {Jenness}, {Kanarek}, {Kendrew},
  {Kern}, {Kerzendorf}, {Khvalko}, {King}, {Kirkby}, {Kulkarni}, {Kumar},
  {Lee}, {Lenz}, {Littlefair}, {Ma}, {Macleod}, {Mastropietro}, {McCully},
  {Montagnac}, {Morris}, {Mueller}, {Mumford}, {Muna}, {Murphy}, {Nelson},
  {Nguyen}, {Ninan}, {N{\"o}the}, {Ogaz}, {Oh}, {Parejko}, {Parley}, {Pascual},
  {Patil}, {Patil}, {Plunkett}, {Prochaska}, {Rastogi}, {Reddy Janga},
  {Sabater}, {Sakurikar}, {Seifert}, {Sherbert}, {Sherwood-Taylor}, {Shih},
  {Sick}, {Silbiger}, {Singanamalla}, {Singer}, {Sladen}, {Sooley},
  {Sornarajah}, {Streicher}, {Teuben}, {Thomas}, {Tremblay}, {Turner},
  {Terr{\'o}n}, {van Kerkwijk}, {de la Vega}, {Watkins}, {Weaver}, {Whitmore},
  {Woillez}, {Zabalza}, \& {Astropy Contributors}}]{Astropy_2018}
{Astropy Collaboration}, {Price-Whelan}, A.~M., {Sip{\H{o}}cz}, B.~M., {et~al.}
  2018, \aj, 156, 123, \dodoi{10.3847/1538-3881/aabc4f}

\bibitem[{Barnes {et~al.}(2017)Barnes, Longmore, Battersby, Bally, Kruijssen,
  Henshaw, \& Walker}]{Barnes_2017}
Barnes, A.~T., Longmore, S.~N., Battersby, C., {et~al.} 2017, \mnras, 469,
  2263, \dodoi{10.1093/mnras/stx941}

\bibitem[{{Benjamin} {et~al.}(2003){Benjamin}, {Churchwell}, {Babler}, {Bania},
  {Clemens}, {Cohen}, {Dickey}, {Indebetouw}, {Jackson}, {Kobulnicky},
  {Lazarian}, {Marston}, {Mathis}, {Meade}, {Seager}, {Stolovy}, {Watson},
  {Whitney}, {Wolff}, \& {Wolfire}}]{Benjamin_2003}
{Benjamin}, R.~A., {Churchwell}, E., {Babler}, B.~L., {et~al.} 2003, \pasp,
  115, 953, \dodoi{10.1086/376696}

\bibitem[{{Bland-Hawthorn} \& {Gerhard}(2016)}]{Bland-Hawthorn_Gerhard_2016}
{Bland-Hawthorn}, J., \& {Gerhard}, O. 2016, \araa, 54, 529,
  \dodoi{10.1146/annurev-astro-081915-023441}

\bibitem[{Callanan {et~al.}(2022, in prep.)Callanan, Longmore, Hatchfield,
  Henshaw, Walker, Battersby, Keto, Barnes, Ginsburg, Kauffmann, Kruijssen, Lu,
  Mills, Pillai, Zhang, Bally, Butterfield, Contreras, Ho, Immer, Johnston,
  Ott, Patel, Tolls, \& et~al.}]{Callanan_2021}
Callanan, D., Longmore, S.~N., Hatchfield, H.~P., {et~al.} 2022, in prep.

\bibitem[{Cram {et~al.}(1996)Cram, Claussen, Beasley, Gray, \&
  Goss}]{Cram_1996}
Cram, L., Claussen, M., Beasley, A., Gray, A., \& Goss, W. 1996, \mnras, 280,
  1110

\bibitem[{{Dahmen} {et~al.}(1998){Dahmen}, {Huttemeister}, {Wilson}, \&
  {Mauersberger}}]{Dahmen_1998}
{Dahmen}, G., {Huttemeister}, S., {Wilson}, T.~L., \& {Mauersberger}, R. 1998,
  \aap, 331, 959.
\newblock \doarXiv{astro-ph/9711117}

\bibitem[{Dame {et~al.}(2001)Dame, Hartmann, \& Thaddeus}]{Dame_2001}
Dame, T.~M., Hartmann, D., \& Thaddeus, P. 2001, \apj, 547, 792,
  \dodoi{10.1086/318388}

\bibitem[{Eden {et~al.}(2020)Eden, Moore, Currie, Rigby, Rosolowsky, Su, Kim,
  Parsons, Morata, Chen, \& et~al.}]{Eden_2020}
Eden, D.~J., Moore, T. J.~T., Currie, M.~J., {et~al.} 2020, \mnras, 498, 5936,
  \dodoi{10.1093/mnras/staa2734}

\bibitem[{{Federrath} {et~al.}(2016){Federrath}, {Rathborne}, {Longmore},
  {Kruijssen}, {Bally}, {Contreras}, {Crocker}, {Garay}, {Jackson}, {Testi}, \&
  {Walsh}}]{Federrath_2016}
{Federrath}, C., {Rathborne}, J.~M., {Longmore}, S.~N., {et~al.} 2016, \apj,
  832, 143, \dodoi{10.3847/0004-637X/832/2/143}

\bibitem[{{Ginsburg} {et~al.}(2016{\natexlab{a}}){Ginsburg}, {Robitaille}, \&
  {Beaumont}}]{ginsburg_pv_2016}
{Ginsburg}, A., {Robitaille}, T., \& {Beaumont}, C. 2016{\natexlab{a}},
  {pvextractor: Position-Velocity Diagram Extractor}.
\newblock \doeprint{1608.010}

\bibitem[{{Ginsburg} {et~al.}(2016{\natexlab{b}}){Ginsburg}, {Henkel}, {Ao},
  {Riquelme}, {Kauffmann}, {Pillai}, {Mills}, {Requena-Torres}, {Immer},
  {Testi}, {Ott}, {Bally}, {Battersby}, {Darling}, {Aalto}, {Stanke},
  {Kendrew}, {Diederik Kruijssen}, {Longmore}, {Dale}, {Guesten}, \&
  {Menten}}]{Ginsburg_2016}
{Ginsburg}, A., {Henkel}, C., {Ao}, Y., {et~al.} 2016{\natexlab{b}}, A\&A, 586,
  A50, \dodoi{10.1051/0004-6361/201526100}

\bibitem[{Ginsburg {et~al.}(2019)Ginsburg, Koch, Robitaille, Beaumont,
  adamginsburg, Sipőcz, ZuHone, Patra, Jones, Lim, Stern, Rosolowsky, Earl,
  de~Val-Borro, jrobbfed, shuokong, Kepley, Sokolov, Badger, Maret, Garrido,
  Booker, \& Tollerud}]{adam_ginsburg_2019_3558614}
Ginsburg, A., Koch, E., Robitaille, T., {et~al.} 2019,
  radio-astro-tools/spectral-cube: Release v0.4.5, v0.4.5,  Zenodo,
  \dodoi{10.5281/zenodo.3558614}

\bibitem[{Gray {et~al.}(1993)Gray, Whiteoak, Cram, \& Goss}]{Gray_1993}
Gray, A., Whiteoak, J., Cram, L., \& Goss, W. 1993, \mnras, 264, 678

\bibitem[{{Hacar} {et~al.}(2022){Hacar}, {Clark}, {Heitsch}, {Kainulainen},
  {Panopoulou}, {Seifried}, \& {Smith}}]{Hacar_2022}
{Hacar}, A., {Clark}, S., {Heitsch}, F., {et~al.} 2022, arXiv e-prints,
  arXiv:2203.09562.
\newblock \doarXiv{2203.09562}

\bibitem[{Hatchfield {et~al.}(2021)Hatchfield, Sormani, Tress, Battersby,
  Smith, Glover, \& Klessen}]{Hatchfield_2021}
Hatchfield, H.~P., Sormani, M.~C., Tress, R.~G., {et~al.} 2021, \apj, 922, 79,
  \dodoi{10.3847/1538-4357/ac1e89}

\bibitem[{Henshaw {et~al.}(2022)Henshaw, Barnes, Battersby, Ginsburg, Sormani,
  \& Walker}]{Henshaw_2022}
Henshaw, J.~D., Barnes, A.~T., Battersby, C., {et~al.} 2022, Star Formation in
  the Central Molecular Zone of the Milky Way,  arXiv,
  \dodoi{10.48550/ARXIV.2203.11223}

\bibitem[{Henshaw {et~al.}(2016)Henshaw, Longmore, Kruijssen, Davies, Bally,
  Barnes, Battersby, Burton, Cunningham, Dale, \& et~al.}]{Henshaw_2016}
Henshaw, J.~D., Longmore, S.~N., Kruijssen, J. M.~D., {et~al.} 2016, \mnras,
  457, 2675, \dodoi{10.1093/mnras/stw121}

\bibitem[{{Humire} {et~al.}(2020){Humire}, {Thiel}, {Henkel}, {Belloche},
  {Loison}, {Pillai}, {Riquelme}, {Wakelam}, {Langer},
  {Hern{\'a}ndez-G{\'o}mez}, {Mauersberger}, \& {Menten}}]{Humire_2020}
{Humire}, P.~K., {Thiel}, V., {Henkel}, C., {et~al.} 2020, \aap, 642, A222,
  \dodoi{10.1051/0004-6361/202038216}

\bibitem[{{Kauffmann} {et~al.}(2017){Kauffmann}, {Pillai}, {Zhang}, {Menten},
  {Goldsmith}, {Lu}, \& {Guzm{\'a}n}}]{Kauffmann_2017}
{Kauffmann}, J., {Pillai}, T., {Zhang}, Q., {et~al.} 2017, \aap, 603, A89,
  \dodoi{10.1051/0004-6361/201628088}

\bibitem[{{Koch} \& {Rosolowsky}(2015)}]{2015MNRAS.452.3435K}
{Koch}, E.~W., \& {Rosolowsky}, E.~W. 2015, \mnras, 452, 3435,
  \dodoi{10.1093/mnras/stv1521}

\bibitem[{Kruijssen \& Longmore(2013)}]{kru_longmore_2013}
Kruijssen, J. M.~D., \& Longmore, S.~N. 2013, \mnras, 435, 2598,
  \dodoi{10.1093/mnras/stt1634}

\bibitem[{Kruijssen {et~al.}(2014)Kruijssen, Longmore, Elmegreen, Murray,
  Bally, Testi, \& Kennicutt}]{Kruijssen_2014}
Kruijssen, J. M.~D., Longmore, S.~N., Elmegreen, B.~G., {et~al.} 2014, \mnras,
  440, 3370, \dodoi{10.1093/mnras/stu494}

\bibitem[{{Liszt}(1992)}]{Liszt_1992}
{Liszt}, H.~S. 1992, \apjs, 82, 495, \dodoi{10.1086/191727}

\bibitem[{{Longmore} {et~al.}(2013){Longmore}, {Bally}, {Testi}, {Purcell},
  {Walsh}, {Bressert}, {Pestalozzi}, {Molinari}, {Ott}, {Cortese}, {Battersby},
  {Murray}, {Lee}, {Kruijssen}, {Schisano}, \& {Elia}}]{Longmore_2013}
{Longmore}, S.~N., {Bally}, J., {Testi}, L., {et~al.} 2013, \mnras, 429, 987,
  \dodoi{10.1093/mnras/sts376}

\bibitem[{{McMullin} {et~al.}(2007){McMullin}, {Waters}, {Schiebel}, {Young},
  \& {Golap}}]{McMullin_2007}
{McMullin}, J.~P., {Waters}, B., {Schiebel}, D., {Young}, W., \& {Golap}, K.
  2007, in Astronomical Society of the Pacific Conference Series, Vol. 376,
  Astronomical Data Analysis Software and Systems XVI, ed. R.~A. {Shaw},
  F.~{Hill}, \& D.~J. {Bell}, 127

\bibitem[{Mills {et~al.}(2018)Mills, Ginsburg, Immer, Barnes, Wiesenfeld,
  Faure, Morris, \& Requena-Torres}]{Mills_2018}
Mills, E. A.~C., Ginsburg, A., Immer, K., {et~al.} 2018, \apj, 868, 7,
  \dodoi{10.3847/1538-4357/aae581}

\bibitem[{Mills \& Morris(2013)}]{Mills_2013}
Mills, E. A.~C., \& Morris, M.~R. 2013, \apj, 772, 105,
  \dodoi{10.1088/0004-637x/772/2/105}

\bibitem[{Molinari {et~al.}(2010)Molinari, Swinyard, Bally, Barlow, Bernard,
  Martin, Moore, Noriega-Crespo, Plume, Testi, \& et~al.}]{Molinari_2010}
Molinari, S., Swinyard, B., Bally, J., {et~al.} 2010, \pasp, 122, 314,
  \dodoi{10.1086/651314}

\bibitem[{{Morris} \& {Serabyn}(1996)}]{morris_serabyn_1996}
{Morris}, M., \& {Serabyn}, E. 1996, \araa, 34, 645,
  \dodoi{10.1146/annurev.astro.34.1.645}

\bibitem[{{Panopoulou} {et~al.}(2022){Panopoulou}, {Clark}, {Hacar}, {Heitsch},
  {Kainulainen}, {Ntormousi}, {Seifried}, \& {Smith}}]{Panopoulou_2022}
{Panopoulou}, G.~V., {Clark}, S.~E., {Hacar}, A., {et~al.} 2022, \aap, 657,
  L13, \dodoi{10.1051/0004-6361/202142281}

\bibitem[{{Salas} {et~al.}(2021){Salas}, {Morris}, \& {Naoz}}]{Salas_2021}
{Salas}, J.~M., {Morris}, M.~R., \& {Naoz}, S. 2021, \aj, 161, 243,
  \dodoi{10.3847/1538-3881/abefd3}

\bibitem[{Shetty {et~al.}(2012)Shetty, Beaumont, Burton, Kelly, \&
  Klessen}]{Shetty_2012}
Shetty, R., Beaumont, C.~N., Burton, M.~G., Kelly, B.~C., \& Klessen, R.~S.
  2012, \mnras, 425, 720

\bibitem[{{Smith} {et~al.}(2020){Smith}, {Tre{\ss}}, {Sormani}, {Glover},
  {Klessen}, {Clark}, {Izquierdo}, {Duarte-Cabral}, \& {Zucker}}]{Smith_2020}
{Smith}, R.~J., {Tre{\ss}}, R.~G., {Sormani}, M.~C., {et~al.} 2020, \mnras,
  492, 1594, \dodoi{10.1093/mnras/stz3328}

\bibitem[{Sormani \& Barnes(2019)}]{Sormani_Barnes_2019}
Sormani, M.~C., \& Barnes, A.~T. 2019, \mnras, 484, 1213,
  \dodoi{10.1093/mnras/stz046}

\bibitem[{{Sormani} {et~al.}(2018){Sormani}, {Tre{\ss}}, {Ridley}, {Glover},
  {Klessen}, {Binney}, {Magorrian}, \& {Smith}}]{Sormani_2018}
{Sormani}, M.~C., {Tre{\ss}}, R.~G., {Ridley}, M., {et~al.} 2018, \mnras, 475,
  2383, \dodoi{10.1093/mnras/stx3258}

\bibitem[{Sormani {et~al.}(2019)Sormani, Treß, Glover, Klessen, Barnes,
  Battersby, Clark, Hatchfield, \& Smith}]{Sormani_2019}
Sormani, M.~C., Treß, R.~G., Glover, S. C.~O., {et~al.} 2019, Monthly Notices
  of the Royal Astronomical Society, 488, 4663–4673,
  \dodoi{10.1093/mnras/stz2054}

\bibitem[{{The GRAVITY Collaboration} {et~al.}(2019){The GRAVITY
  Collaboration}, {Abuter, R.}, {Amorim, A.}, {Baub\"ock, M.}, {Berger, J. P.},
  {Bonnet, H.}, {Brandner, W.}, {Cl\'enet, Y.}, {Coud\'e du Foresto, V.}, {de
  Zeeuw, P. T.}, {Dexter, J.}, {Duvert, G.}, {Eckart, A.}, {Eisenhauer, F.},
  {F\"orster Schreiber, N. M.}, {Garcia, P.}, {Gao, F.}, {Gendron, E.},
  {Genzel, R.}, {Gerhard, O.}, {Gillessen, S.}, {Habibi, M.}, {Haubois, X.},
  {Henning, T.}, {Hippler, S.}, {Horrobin, M.}, {Jim\'enez-Rosales, A.},
  {Jocou, L.}, {Kervella, P.}, {Lacour, S.}, {Lapeyr\`ere, V.}, {Le Bouquin,
  J.-B.}, {L\'ena, P.}, {Ott, T.}, {Paumard, T.}, {Perraut, K.}, {Perrin, G.},
  {Pfuhl, O.}, {Rabien, S.}, {Rodriguez Coira, G.}, {Rousset, G.},
  {Scheithauer, S.}, {Sternberg, A.}, {Straub, O.}, {Straubmeier, C.}, {Sturm,
  E.}, {Tacconi, L. J.}, {Vincent, F.}, {von Fellenberg, S.}, {Waisberg, I.},
  {Widmann, F.}, {Wieprecht, E.}, {Wiezorrek, E.}, {Woillez, J.}, \& {Yazici,
  S.}}]{grav_collab_2019}
{The GRAVITY Collaboration}, {Abuter, R.}, {Amorim, A.}, {et~al.} 2019, A\&A,
  625, L10, \dodoi{10.1051/0004-6361/201935656}

\bibitem[{Tress {et~al.}(2020)Tress, Sormani, Glover, Klessen, Battersby,
  Clark, Hatchfield, \& Smith}]{Tress_2020}
Tress, R.~G., Sormani, M.~C., Glover, S. C.~O., {et~al.} 2020, \mnras, 499,
  4455, \dodoi{10.1093/mnras/staa3120}

\bibitem[{Wilson \& Rood(1994)}]{wilson_rood_1994}
Wilson, T.~L., \& Rood, R.~T. 1994, \araa, 32, 191,
  \dodoi{10.1146/annurev.aa.32.090194.001203}

\bibitem[{Zucker \& Chen(2018)}]{Zucker_2018}
Zucker, C., \& Chen, H. H.-H. 2018, \apj, 864, 152,
  \dodoi{10.3847/1538-4357/aad3b5}

\end{thebibliography}
\bibliographystyle{aasjournal}



\end{document}